\documentclass[twocolumn]{aastex63}
\usepackage{amsmath}

\newcommand{\vdiff}{v_{\rm diff}}
\newcommand{\tdiff}{t_{\rm diff}}
\newcommand{\vdiffn}{v_{\rm diff,num}}
\newcommand{\lcool}{\lambda_{\rm F, turb}}
\newcommand{\tcool}{t_{\rm cool}}

\newcommand{\tcoolmin}{t_{\rm cool, min}}

\newcommand{\edot}{\dot{E}_{\rm cool}}
\newcommand{\edotc}{\dot{\varepsilon}_{\rm cool}}
\newcommand{\edotcmax}{\dot{\varepsilon}_{\rm max}}
\newcommand{\edoth}{\dot{\varepsilon}_{\rm heat}}
\newcommand{\tsh}{t_{\rm sh}}
\newcommand{\Lbox}{L_{\rm box}}
\newcommand{\Lint}{L_{\rm int}}
\newcommand{\vtL}{v_t(\Lint)}
\newcommand{\vtl}{v_t(\ell)}
\newcommand{\teddy}{t_{\rm eddy}}

\newcommand{\Da}{{\rm Da}}
\newcommand{\Dasim}{{\rm Da}_{\rm sim}}

\newcommand{\rhobar}{\rho_0}
\newcommand{\vrel}{v_{\rm rel}}
\newcommand{\mach}{\mathcal{M}}
\newcommand{\Pbar}{P_0}
\newcommand{\cs}{c_s}
\newcommand{\cshot}{c_{s,{\rm hot}}}

\newcommand{\cspk}{c_{s,{\rm pk}}}
\newcommand{\Tcold}{T_{\rm cold}}
\newcommand{\Tpk}{T_{\rm pk}}
\newcommand{\Thot}{T_{\rm hot}}

\newcommand{\Tlo}{T_{\rm lo}}
\newcommand{\Thi}{T_{\rm hi}}
\newcommand{\Tlim}{T_{\rm lim}}
\newcommand{\fint}{f_{\rm int}}
\newcommand{\bhi}{\beta_{\rm hi}}
\newcommand{\blo}{\beta_{\rm lo}}
\newcommand{\ch}{c_{\rm heat}}
\newcommand{\ah}{\alpha_{\rm heat}}

\newcommand{\Aint}{A_{\rm int}}
\newcommand{\nhat}{\mathbf{\hat{n}}}
\newcommand{\vvec}{\mathbf{v}}
\newcommand{\vinloc}{v_{\rm in}}
\newcommand{\fwindow}{f_w}
\newcommand{\vbulk}{v_{\rm bulk}}
\newcommand{\anum}{\alpha_{\rm num}}

\newcommand{\Vmix}{V_{\rm mix}}

\newcommand{\vzpk}{\left\langle v_z\right\rangle_{\rm pk}}

\newcommand{\dx}{\Delta x}
\newcommand{\nres}{N_{\rm res}}

\newcommand{\meanP}{\left\langle P \right\rangle}
\newcommand{\Rzz}{\mathcal{R}_{zz}}

\newcommand{\ak}{\texttt{AthenaK}\ }
\newcommand{\app}{\texttt{Athena++}\ }

\newcommand{\paperii}{Paper 2\,}

\received{XXX}
\revised{XXX}
\accepted{XXX}

\submitjournal{ApJ}

\shorttitle{This is not a Turbulent Mixing Layer}
\shortauthors{Lancaster et al.}
\begin{document}

\title{Ceci n'est pas une Couche de M\'{e}lange:\\
The Meaning of Resolved Turbulent Radiative Mixing}

\correspondingauthor{Lachlan Lancaster}
\email{llancaster@flatironinstitute.org}

\author[0000-0002-0041-4356]{Lachlan Lancaster}
\thanks{Simons Fellow}
\affiliation{Department of Astronomy, Columbia University,  550 W 120th St, New York, NY 10025, USA}
\affiliation{Center for Computational Astrophysics, Flatiron Institute, 162 5th Avenue, New York, NY 10010, USA}

\author[0000-0002-1600-7552]{Rajsekhar Mohapatra}
\affiliation{Department of Astrophysical Sciences, Princeton University, Princeton, NJ 08544, USA}

\author[0000-0003-3806-8548]{Drummond B. Fielding}
\affiliation{Department of Physics, New York University, 726 Broadway, New York, NY, 10003, USA}

\author[0000-0003-2630-9228]{Greg L. Bryan}
\affiliation{Department of Astronomy, Columbia University,  550 W 120th St, New York, NY 10025, USA}
\affiliation{Center for Computational Astrophysics, Flatiron Institute, 162 5th Avenue, New York, NY 10010, USA}

\begin{abstract}
Turbulent Radiative Mixing Layers (TRMLs) are of fundamental importance to the transport of energy and momentum in multi-phase, astrophysical fluids. We use measurements of the ``micro'' and ``macro'' properties of these layers in high-resolution \ak simulations to investigate when their properties can be considered \textit{well}-resolved. In particular, we demonstrate that the previously noticed resolution independence of total cooling, $\edot$, in these simulations is due to a remarkable, and perhaps fortuitous, cancellation of the countervailing effects of numerical dissipation and numerical viscosity. This calls into question the degree to which we can trust the results of these experiments, as there is no physical picture that explains this cancellation. We also demonstrate that in order to correctly resolve the phase structure in these layers, important for accurate predictions of their observable properties, one must resolve the scale on which turbulent diffusion acts on time-scales comparable to the cooling time. This ``turbulent Field length'', $\lcool$, is where the eddy turnover time is equal to the cooling time ($\teddy(\lcool) = \tcool$). We demonstrate that resolving this scale results in converged phase-structure and spatially resolved transitions in the gas phases.
\end{abstract}


\section{Introduction}
\label{sec:intro}

Astrophysical fluid dynamics is almost always in the regime where the size of the system of interest, $L$, is much larger than the `diffusive' scales on which fluid elements mix together in a microphysical manner. In the case of diffusion of momentum, this means astrophysics involves high Reynolds number (turbulent) flows. As the sources of energy in astrophysics are usually highly structured in time and space, and diffusion of that energy through the system is relatively slow (the diffusive scales are small) astrophysics is also inherently the study of `multi-phase' fluids where high and low specific-energy gas live in close proximity. Crucially, turbulence induces an `effective' diffusivity in astrophysical gas by drastically enhancing the surface area for micro-physical interaction (stretching-enhanced diffusion) \citep{MoninYaglomBook,Frisch95,PopeTF}. This effective diffusivity is naturally much larger than the micro-physical diffusivity in the system, so that turbulence is the dominant mechanism allowing for energy to be communicated between the phases of astrophysical fluids.

It is not surprising, then, that interface regions between fluid phases where turbulence acts to mix the phases and some subsequent reaction or thermal instability occurs, are ubiquitous in astrophysics. These turbulent radiative mixing layers (TRMLs), can be found in the neutral phase of the interstellar-medium \citep{KK13,JenningsLi21}, at the edges of supernova and wind-blown bubbles in star forming regions \citep{Weaver77,ElBadry19,Lancaster21a,Lancaster21b,Pittard22Winds,Lancaster24a}, at the boundaries of warm clouds entrained in hot galactic winds \citep{Armillotta17,GronkeOh18,GronkeOh20a,LiHopkins20,Sparre20,Abruzzo22,Abruzzo24,ChenOh24,FB22,WarrenSchneider25}, at the interaction of accreting cosmic filaments with the circum-galactic medium (CGM) \citep{Mandelker20}, or thermally unstable gas within the CGM and the Intracluster Medium (ICM) \citep{Mohapatra23,Mohapatra25}, in the tails of ram-pressure stripped galaxies \citep{Simons2020,TonnesenBryan21}, and similar physics plays a role in turbulent nuclear deflagrations \citep{Roepke07}. Understanding these systems in the physical world relies on spectral diagnostics: emission and absorption lines \citep[e.g.][]{JenkinsMeloy74,Werk14,Zahedy19}. These lines are very sensitive to the exact temperature/density distributions in these mixing layers, meaning that accurate theoretical models of this structure are essential.

Given the wide applicability of this physics in astrophysical fluid dynamics, many previous authors have attempted to understand what aspects of the turbulent flow impact the communication of energy, momentum, and mass between the phases \citep{Ji19,FieldingFractal20,Tan21,ZhaoBai23,Das24,TMG25}. These works have generally found that the ``zeroth order'' results of these calculations, such as the total energy and momentum transfer per unit time, are remarkably independent of how well resolved the turbulent layers are. In this work we demonstrate that this resolution independence is enabled by a possibly serendipitous cancellation of the effects of numerical diffusivity and numerical viscosity (this perhaps mimics real physics). We also demonstrate, however, that the accurate representation of the ``first order'' properties of the layer such as its phase structure (and consequently its line emission) has a more stringent resolution requirement.

\begin{figure*}
    \centering
    \includegraphics[width=\textwidth]{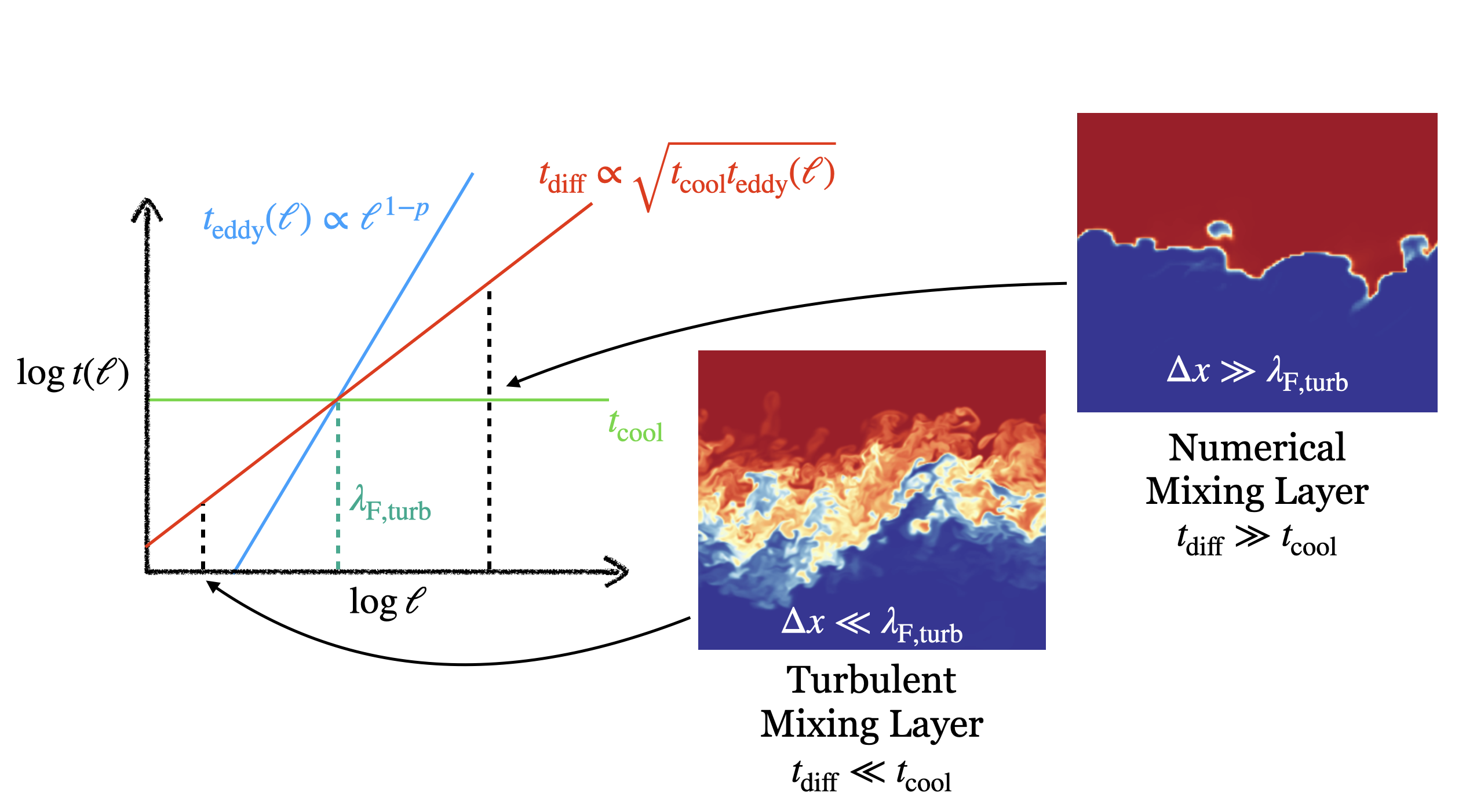}
    \caption{At left we show a schematic representation of several timescales as a function of physical scale, $\ell$: the cooling time, $\tcool$, the eddy turnover time, $\teddy(\ell)$ (\autoref{eq:teddy_def}), and the turbulent diffusive time, $\tdiff$ (\autoref{eq:tdiff_geo_mean}), which is their geometric mean, all as a function of scale. The intersection of $\teddy$ and $\tcool$ defines the critical scale, $\lcool$, at which the strength of turbulent diffusion and cooling balance. When $\tdiff \gg \tcool$ ($\ell \gg \lcool$) diffusion is not able to act on the timescales on which cooling evolves. Simulations which are only able to represent these scales (like that on the far right) do not accurately represent turbulent diffusion, but rather numerically mix gas into the cooling layer, where cooling acts to change its internal energy rapidly. Higher resolution simulations (like that in the center of the figure) resolve the scales on which diffusion is able to act on timescales shorter than the cooling time, this is the hallmark of a true mixing layer.}
    \label{fig:resolution_schematic}
\end{figure*}

The paper is structured as follows. In \autoref{sec:theory} we review the theory of turbulent mixing in a general language that informs the discussion in the remainder of the paper. In \autoref{sec:methods} we describe the numerical simulations we present here and in \autoref{sec:measurement} we detail the measurements we perform on them. In \autoref{sec:results} we demonstrate that resolving a scale we term the turbulent-Field length\footnote{This length will be the scale on which turbulent diffusion is balanced by cooling, and is named in analogy to the Field length \citep{Field65} on which micro-physical thermal conduction balances cooling.}, $\lcool$, leads to converged phase structure in the layer. We emphasize (\autoref{fig:resolution_schematic}) that simulations which do not resolve this scale are not representative of \textit{true} mixing layers, but are rather ``numerical mixing layers.'' We discuss the implications of these results in the context of past work in \autoref{sec:discussion}. In a companion paper, which we will refer to as \paperii, we discuss the behavior of these mixing layers in the regime where cooling becomes very fast in comparison to turbulent mixing.

\section{Theory of Radiative Mixing Layers}
\label{sec:theory}

In this work we are interested in how energy is lost when two thermally stable\footnote{Or at least ``quasi''-thermally stable in that the time scales for thermal instability are much longer than the dynamical times of the system.} gas phases are mixed together through turbulence to create an unstable phase in which some reaction takes place. We describe the turbulence by its velocity scale, $\vtl$, as a function of physical scale $\ell$. We will assume that $\vtl$ is given by the square-root of the 2nd-order structure function\footnote{In this work, when referring to $i^{\rm th}$ structure function of a field, $f$,  ${\rm SF}_i(f)$, we will be referring to the $i^{\rm th}$-root of the traditionally defined structure function, so that structure functions of all orders have the same dimensions as the underlying field, $f$.} of the turbulent velocity field given by
\begin{equation}
    \label{eq:vsf2_def}
    {\rm SF}_2(\vvec;\ell) = \left[\frac{1}{V}\int_V \left\langle|\vvec_t(\mathbf{r}) - \vvec_t(\mathbf{r} + \mathbf{l})|^2 \right\rangle_{|\mathbf{l}| = \ell} d\mathbf{r} \right]^{1/2}
\end{equation}
where the angle brackets indicate an average over all spatial deviations $\mathbf{l}$ whose norms are equal to $\ell$. In other words, this is the root-mean-square velocity difference on a scale $\ell$ in volume $V$.

We will assume that the turbulence has some outer `energy-containing' or `integral' scale, $\Lint$, such that most turbulent kinetic energy is contained in fluctuations at this scale. We define the eddy turnover time of the turbulence as 
\begin{equation}
    \label{eq:teddy_def}
    \teddy(\ell) \equiv \frac{\ell}{\vtl} \, .
\end{equation}

We then suppose that the cooling reaction that occurs when the two phases are mixed has some characteristic associated timescale, $\tcool$. The exact definition of this time-scale varies across the literature and to some extent depends on the details of the associated cooling function. In this section we will only discuss it abstractly, but in the remainder of the work we will use the minimum isobaric cooling time across the mixed gas, $\tcoolmin$ (see \autoref{subsec:cool_func}).

One key dimensionless quantity is the Damk\"{o}hler number \citep{damkohler40,kuo12,PoinsotBook,Tan21}
\begin{equation}
    \label{eq:Da_def}
    \Da(\ell) \equiv \frac{\teddy(\ell)}{\tcool} \, .
\end{equation}
Here, we have intentionally written $\Da$ as a function of scale, $\ell$, though it is generally defined for a system by $\Da = \Da(\Lint)$ and when we mention $\Da$ by itself we will mean it in this manner. Assuming Kolmogorov turbulence $\teddy\propto \ell^{2/3}$, implying that mixing becomes increasingly dominant to cooling at small scales.

As energy is lost through the reaction, this loss is balanced by an inflow of higher specific entropy gas to the layer. If this inflow is mediated by some thermal diffusivity $\alpha_T$ (with dimensions of length squared over time) then one can show\footnote{This can be done either through solving a time-steady one-dimensional energy equation including the thermal diffusivity and appropriate reaction terms or simple dimensional analysis, as is shown in the cited works.} that the velocity at which gas is brought in to the layer mediated by this diffusivity is \citep{Field65,ZDPN69,KK13,Tan21}
\begin{equation}
    \label{eq:vdiff_scale}
    \vdiff \approx \sqrt{\frac{\alpha_T}{\tcool}} \, .
\end{equation}

In order to maintain energy balance, the cooling rate within the layer must be balanced by the enthalpy flux into the layer so that we have
\begin{equation}
    \label{eq:edot_generic}
    \edot = \frac{\gamma}{\gamma -1}P \vdiff \Aint
\end{equation}
where $\Aint$ is the area of the interface set up between the two phases. \autoref{eq:edot_generic} writes the enthalpy flux into the layer, considering the full geometry of the layer on the smallest scales that are not made laminar by the diffusivity.

If we imagine that some small portion of a mixing layer, occupying a volume $\ell^3$, cools on a timescale $\tcool$, the time it takes for energy to be re-supplied by the diffusive transport through the side of this volume (area $\ell^2$) is
\begin{equation}
    \label{eq:tdiff_def}
    \tdiff (\ell) = \frac{\ell}{v_{\rm diff}} \approx \ell\sqrt{\frac{\tcool}{\alpha_T}} \, .
\end{equation}

If we then imagine that the diffusivity is mediated by some ``effective" turbulent diffusion on a scale $\ell$, $\alpha_T= \vtl \ell$, with a turbulent velocity given by \autoref{eq:vsf2_def} then the diffusive timescale becomes
\begin{equation}
    \label{eq:tdiff_geo_mean}
    \tdiff(\ell) \approx \sqrt{\teddy(\ell) \tcool} \, .
\end{equation}
That is, the geometric mean of the cooling time and the eddy turnover time. On the left hand side of \autoref{fig:resolution_schematic} we show a schematic of how each of these timescales vary as a function of length scale (assuming that $\vtl\propto \ell^{p}$ with $p<1$).

In this scenario, the population of gas at intermediate temperatures between the two phases is determined by the competition of the effective turbulent diffusion and cooling across the layer. If a simulation cannot represent the scales on which these processes are faithfully represented, with $\tdiff \ll \tcool$, it will not correctly reproduce the phase transition. In particular, if the parameters of a given numerical simulation are such that $\teddy(\ell) \gg \tcool$ on resolved scales (right hand side of \autoref{fig:resolution_schematic}), numerical diffusion\footnote{To be clear, even in our simulations which resolve $\tdiff \ll \tcool$ mixing of the two phases is still fundamentally caused by numerical diffusion, as we do not include explicit diffusion, which should mix the phases on small scales.} will act to mix the two phases, at which point cooling will act to quickly remove energy from the mixed gas, keeping the interface thin, with a thickness on the order of the resolution, $\dx$. In the opposite limit, when the simulation is able to resolve scales where $\teddy (\ell) \ll \tcool$ (middle of \autoref{fig:resolution_schematic}), turbulent diffusion is able to wacompete on fair footing with cooling and faithfully represent the way in which the surface area is enhanced on small scales, which (combined with physical diffusivity) acts to smooth out the temperature distribution of the interface.

In order to resolve this transition then, we must resolve the scales on which turbulent diffusion evolves on a time-scale comparable to the cooling time. We therefore define the \textit{turbulent Field length}
\begin{equation}
    \label{eq:lcool_def}
    \teddy(\lcool) = \tcool
\end{equation}
in analogy with the Field length on which thermal conduction and cooling balance \citep{Field65,BegelmanMcKee90,KoyamaInutsuka04}, this is the scale on which cooling is balanced by an effective turbulent diffusivity on that scale: $\alpha_T = v_t(\lcool)\lcool$. The balancing of $\alpha_T$ against cooling is somewhat tautological as we have essentially included cooling in our definition of $\tdiff$ in \autoref{eq:tdiff_def}. We are essentially saying here that turbulent diffusion can only act as an effective diffusivity on scales $\ell \leq \lcool$, as on scales larger than this cooling acts to sharpen the interface faster than turbulence can act to diffuse across it. We then define $\lcool$ as a ``Field Length'' in balancing cooling against turbulent diffusion at the largest scale that it can be considered to effectively behave diffusively, which is $\lcool$ itself. This scale is the same scale referred to as $w$ in \citet{FieldingFractal20} and $\ell_{\rm cool}$ in \citet{Lancaster24a} and, considering \autoref{eq:Da_def}, is the scale on which $\Da(\lcool) = 1$. Indeed, this scale may be more appropriately thought of as a ``cooling length'' (traditionally $\cs\tcool$) where we replace the sound speed with a specific turbulent velocity (the turbulent velocity at scale $\ell$ where $\teddy(\ell)$ is equal to $\tcool$). Below this scale, the fluid is essentially isothermal due to rapid mixing. A truly resolved mixing layer will have a ``width'' of the temperature transition that is on the order of $\lcool\gg \dx$ (hence the use of $w$ in \citet{FieldingFractal20}).

We have made an argument for what it means to resolve a mixing layer in terms of what resolution is needed in order to correctly resolve the phase transition across the layer. We will see in \autoref{subsec:phase} that this is indeed the case. However, the unresolved layer with $\dx \gg \lcool$ may still correctly represent the total amount of diffusion across the layer, as has indeed been observed in past studies \citep{Ji19,FieldingFractal20,Tan21,Lancaster21b}. In \autoref{sec:results}, we will explain how this occurs.

\begin{figure}
    \centering
    \includegraphics{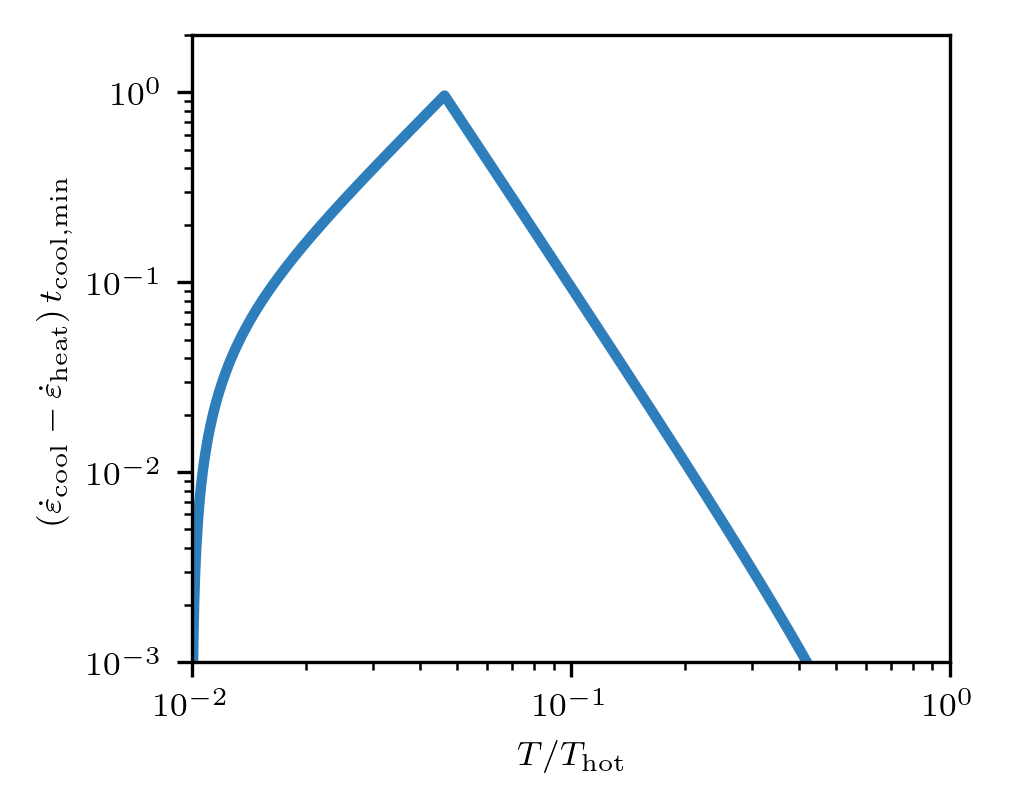}
    \caption{We show the net cooling function, $\edotc-\edoth$, used in the simulations presented in this work with $\chi =10^2$ normalized by the minimum cooling rate, which varies between simulations based on the parameter $\xi$ (see text). The above plot assumes the gas is isobaric at the background pressure, $\Pbar$.}
    \label{fig:cf}
\end{figure}

\begin{figure*}
    \centering
    \includegraphics[width=\textwidth]{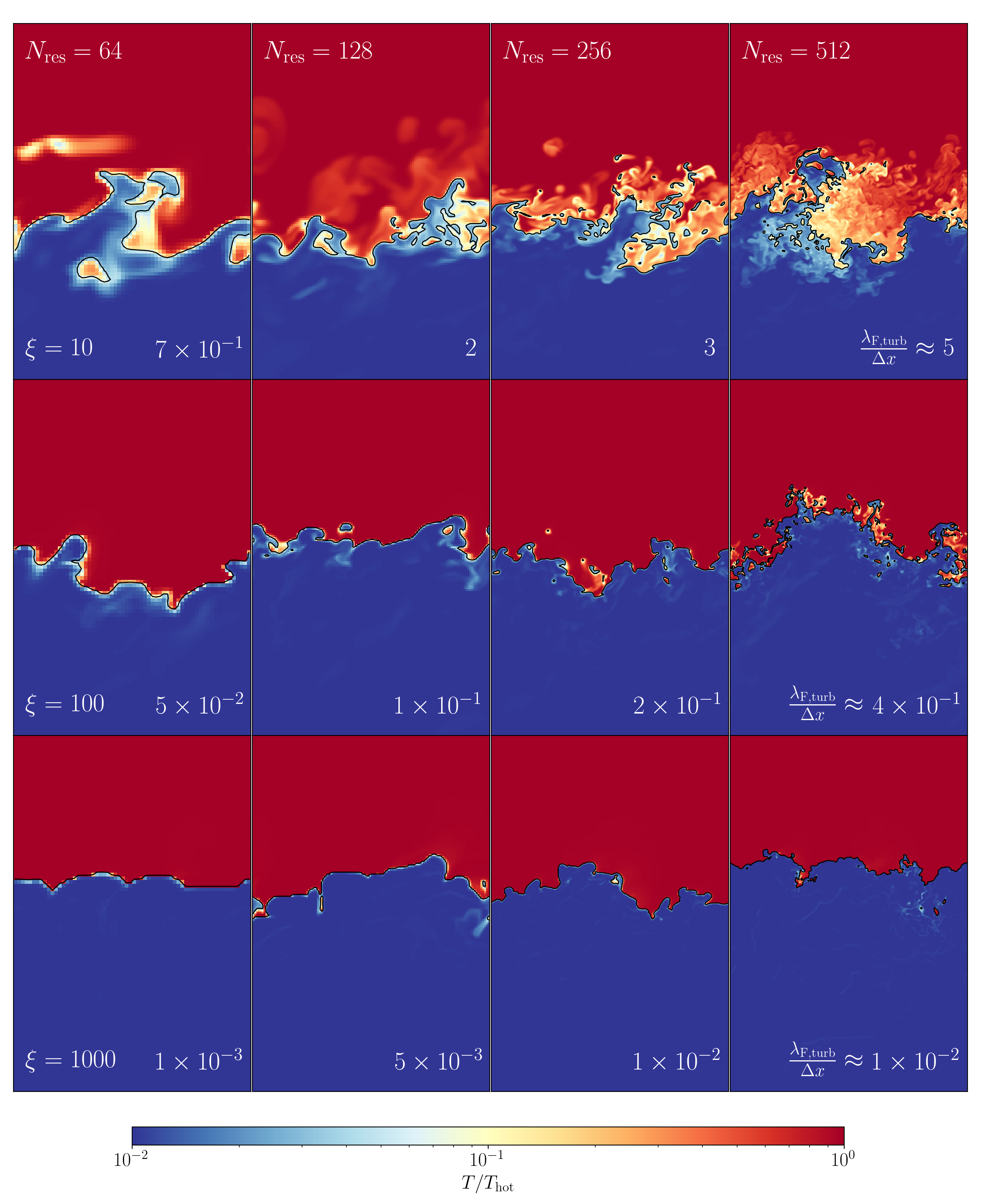}
    \caption{We show slices of the gas temperature through the center of the domain at $t = 22.5\,\tsh$ in our $\mach = 1/2$ suite of simulations at varying resolution ($\nres = 64,\, 128,\, 256,\,\&\, 512$ left to right) and $\xi= 10,\, 100,\, 1000$ (top to bottom). The numbers in the bottom right of each panel give the approximate value of $\lcool/\dx$ for each simulation. From right to left and top to bottom this ratio changes by approximately a factor of two in each panel. Black lines indicate the iso-contours of $T = \Tpk$. Simulations at the right of the top-row are considered to be marginally resolved by the standards of this work.}
    \label{fig:multi_panel}
\end{figure*}

\section{Numerical Experiment Design}
\label{sec:methods}

We employ the GPU-accelerated, magneto-hydrodynamical (MHD) code \ak \citep{athenak} to simulate the Euler Equations of ideal hydrodynamics with a cooling source term applied to the energy equation, detailed in \autoref{subsec:cool_func}. \ak solves the Euler Equations using a finite-volume scheme. We employ the piecewise linear method (PLM) for 2nd-order spatial reconstruction of the primitive variables to cell interfaces. Fluxes at these interfaces are computed using the HLLC Riemann solver \citep{HLLC94} and the conservative variables are updated in time using these fluxes along with a 2nd-order Runge-Kutta (RK2) time integration scheme. Further details on the numerical methods can be found in the \app methods paper of \citet{athenapp}, the original \texttt{Athena} methods paper of \citet{Stone08_Athena}, and references therein.

\subsection{Initial Conditions}
\label{subsec:ics}

We simulate a box of dimensions $(L_x,\, L_y,\, L_z) = (\Lbox,\, \Lbox,\, 1.5\, \Lbox)$\footnote{Our simulations with $\xi = 1$ have $L_z =3\,\Lbox$ to avoid the layer filling the domain. $z$ is understood to vary between ($-\Lbox$, $2\,\Lbox$ in this case).} with periodic boundary conditions at the $x$ and $y$ interfaces and an outflow boundary condition at the bottom $z$ boundary (top-$z$ boundary condition given below). $\Lbox=1$ throughout this work, though we will use $\Lbox$ abstractly below to refer to the side-length of the box. We will refer to the $z$-coordinate as varying within $z = (-\Lbox/2,\, \Lbox)$.

We initialize the primitive variables within the domain with two constant states which are transitioned between at $z=0$ using a tanh profile of thickness $\Delta x/2$, with $\dx \equiv \Lbox/\nres$ the physical resolution and $\nres$ the number of resolution elements in the $x$ and $y$ directions. We use constant spatial resolution everywhere, so the $z$-direction is resolved by $1.5\times\nres$ resolution elements \textbf{($3\,\nres$ for $\xi=1$)}. We refer to the state at $z>0$ ($z<0$) as the `hot' (`cold') gas. The primitive variables in these states are given by
\begin{equation}
    \label{eq:ics}
    \begin{pmatrix}
        \rho_{\rm hot} \\ v_{x,{\rm hot}} \\ v_{y,{\rm hot}} \\ v_{z,{\rm hot}} \\ P_{\rm hot}
    \end{pmatrix}
    = 
    \begin{pmatrix}
        \rhobar \\ \vrel/2 \\ 0 \\ 0 \\ \Pbar
    \end{pmatrix}
    \,\,\, , \,\,\,
    \begin{pmatrix}
        \rho_{\rm cold} \\ v_{x,{\rm cold}} \\ v_{y,{\rm cold}} \\ v_{z,{\rm cold}} \\ P_{\rm cold}
    \end{pmatrix}
    = 
    \begin{pmatrix}
        \chi \rhobar \\ -\vrel/2 \\ 0 \\ 0 \\ \Pbar
    \end{pmatrix} \, .
\end{equation}
$\rhobar = \Pbar = 1$ throughout this work but, as with $\Lbox$, will be referred to in the abstract for generality. $\vrel$ is the relative velocity between the two layers and the density contrast, $\chi$. This density contrast is representative of the density contrast between cool clouds ($\sim 10^4\, {\rm K}$) and the surrounding hot wind ($\sim 10^6\,{\rm K}$) in a galactic wind. We will make the simplification $T = P/\rho$ so that temperature in our study has units of velocity squared. The adiabatic sound speed in the hot medium is then $\cshot = \sqrt{\gamma \Pbar/\rhobar}$ where we take $\gamma = 5/3$, as appropriate for a monatomic gas. The temperatures of the two phases are then given by $\Thot = \Pbar/\rhobar$ and $\Tcold = \Pbar/\chi\rhobar = \Thot/\chi$. The top-$z$ boundary condition is given by fixing $\mathbf{U} = \mathbf{U}_{\rm hot}$ for all conserved variables except $\rho v_z$ which employs a zero-gradient condition, to allow for an arbitrary inflow.

\subsection{Cooling Function}
\label{subsec:cool_func}

We employ a source term to the energy equation in order to represent thermal heating and cooling, implemented in an operator-split fashion. Our source term is very similar to that of \citet{CFB23}: it guarantees two thermally stable phases at $\Thot$ and $\Tcold$ and has a cooling rate which peaks at an intermediate temperature $\Tpk \equiv \left(\Tcold^2 \Thot \right)^{1/3}$ with a power-law fall-off in cooling rate on either side of this peak, as shown in \autoref{fig:cf}. We employ this simplified cooling function in order to investigate a minimally complicated test scenario for TRMLs, so that the complications of complex features in realistic cooling functions do not obscure the fundamental fluid physics we are trying to investigate.

Specifically, the cooling rate density is taken to be a function of pressure and temperature as
\begin{equation}
    \label{eq:edot_cool}
    \edotc(P,T) = \edotcmax(P) \left(\frac{T}{\Tpk} \right)^{-\beta(T)}
\end{equation}
where 
\begin{equation}
    \label{eq:edc_max}
    \edotcmax(P) \equiv \frac{\Pbar/(\gamma - 1)}{\tcoolmin}\left(\frac{P}{\Pbar} \right)^2
\end{equation}
is the maximum cooling rate and $\tcoolmin$ is the minimum cooling time, both at a given pressure. We define
\begin{equation}
    \beta(T) = 
    \begin{cases} 
      \blo & T < \Tpk \\
      \bhi & T > \Tpk
   \end{cases} \, ,
\end{equation}
with $\blo < 0$, $\bhi > 0$ in general. We choose $(\blo,\, \bhi) = (-2,\, 3) \times 2/\log_{10}(\chi)$. The $\chi$ dependence in $\blo$ and $\bhi$ is included so that $\edotc(\Tcold)$ and $\edotc(\Thot)$ are identical across simulations with the same value of $\xi$.

The heating rate density is given by
\begin{equation}
    \label{eq:edot_heat}
    \edoth = \ch\edotcmax(P) f_h(T)
\end{equation}
where
\begin{equation}
    \ch = \left(\frac{\Tcold}{\Tpk} \right)^{(\bhi-\blo) \left(1 + \frac{\log(\Tcold/\Tpk)}{\log(\chi)} \right)}
\end{equation}
and
\begin{equation}
    \label{eq:fhT}
    f_h(T) = 
    \begin{cases}
        \left(\frac{T}{\Tpk} \right)^{\ah} & T<\Tlim\\
        \left(\frac{\Tlim}{\Tpk} \right)^{\ah} \left(\frac{T}{\Tlim} \right)^{-\bhi-1/2} & T > \Tlim
    \end{cases}
\end{equation}
with
\begin{equation}
    \ah = \left(\blo - \bhi \right)\left(\frac{\log(\Tcold/\Tpk)}{\log(\chi)} \right) - \bhi
\end{equation}
and $\Tlim \equiv 1.05\times\Thot$ intended as a `limiting temperature.' We refer to it this way because the second clause in \autoref{eq:fhT} is meant to decrease the heating rate more rapidly as one goes to temperatures above $\Tlim$ so that eventually cooling is dominant again. While this is included in principle to prevent shocks in the hot gas from creating too high temperature, it is never an important term in practice. To reiterate, $\ch$ and $\ah$ are generally chosen in order to assure $\edotc(\Pbar,\Tcold) = \edoth(\Pbar, \Tcold)$ and $\edotc(\Pbar,\Thot) = \edoth(\Pbar, \Thot)$, so that we have two thermally stable phases.

The quadratic dependence on $P$ in \autoref{eq:edc_max} is included to mimic the typical density squared dependence of collisional cooling terms in astrophysical gases, though this analogy is limited given that this dependence is also applied to the heating terms, which typically have a linear dependence on density \citep{Drainebook}. Regardless, our choice of cooling function is idealized and only meant to be representative of a generalized physics problem, the inclusion or exclusion of this term is not expected to impact the main conclusions of this work. The total source term to the energy equation is given by $\edoth(P,T)  - \edotc(P,T)$.%

\subsection{Frame Tracking}
\label{subsec:frame_track}

In order to make sure that the mixing layer remains within the bounds of our simulation domain we employ Galilean velocity boosts in the $z$-direction to track the front. Specifically, every $10$ simulation time steps we measure the mean $v_z$ of peak cooling gas
\begin{equation}
    \vzpk = {\rm avg}\left(\left\lbrace v_z(\mathbf{x}) : 1/\fwindow<T(\mathbf{x})/\Tpk < \fwindow \right\rbrace\right) 
    \, ,
\end{equation}
with $\fwindow=10^{0.1}$, and apply a Galilean boost of $-\vzpk$ to the entire domain. If this velocity shift would bring the layer too close to the upper ($0.8\,\Lbox$) or lower ($-0.4\,\Lbox$) boundary of the domain, we apply a velocity shift away from the boundary that is proportional to the distance between the layer and the boundary. We also limit this velocity shift to $|\vzpk| < \vrel/100$.

\subsection{Parameter Study}
\label{subsec:suite_deets}

Here we define the main parameters relevant to a mixing layer simulation and outline the range of those parameters explored in this work. We first define
\begin{equation}
    \label{eq:xi_def}
    \xi \equiv \frac{\tsh}{\tcoolmin}
\end{equation}
the ratio of the shear time
\begin{equation}
    \label{eq:tshear}
    \tsh \equiv \frac{\Lbox}{\vrel}
\end{equation}
to the minimum cooling time. We choose $\xi$ as the key parameter describing cooling in our simulations, as in \citet{FieldingFractal20}. In \paperii we discuss $\xi$ in relation to $\Da$ as well as other definitions for $\Da$ used in previous works.

The other key parameters of the simulations are the relative shear velocity, $\vrel$ which we specify by a choice of the Mach number of this shear in the hot gas
\begin{equation}
    \label{eq:mach_def}
    \mach \equiv \frac{\vrel}{\cshot} \, .
\end{equation}
and the density contrast
\begin{equation}
    \label{eq:chi_def}
    \chi \equiv \frac{\rho_{\rm cold}}{\rho_{\rm hot}} = \frac{\Thot}{\Tcold}
\end{equation}

Our suite consists of two choices for the Mach number ($\mach = 1/2$ and $\mach=1/8$). As super-sonic TRMLs will result in shocks that make the gas immediately adjacent to the mixing layer subsonic, the cases investigated here may be representative of slightly more general scenarios (see \citet{YangJi23} for a more detailed analysis of the high $\mach$ case). Our whole suite consists of simulations with $\chi = 30,\, 10^2\, \&\, 10^3$. We also run a series of simulations at $\chi=10$ some of which are used in \autoref{app:res_dep}. Due to the resolution dependence discussed in that appendix we do not further analyze the $\chi =10$ simulations here or in \paperii).

As $\Lbox$ is fixed, $\tsh$ is fixed given $\vrel$. We then specify $\tcoolmin$ (and therefore the cooling function overall) by a choice of $\xi$. We run simulations at $\xi = 1,\, 3,\, 10,\, 30,\, 100,\, 300,\, 1000$.

We run each simulation for $t_{\rm sim} = 30\, \tsh$ and perform all our analysis on times $t>10\tsh$, which is when the mixing layer reaches a steady state. For each simulation we output 100 snapshots, equally spaced in time. In \autoref{fig:multi_panel} we display slices of the gas temperature across a range of resolution and $\xi$ values for the $\chi = 10^2$, $\mach = 1/2$ suite of simulations.

\section{Measurement Techniques}
\label{sec:measurement}

In this section we provide details on how all quantities of interest are measured in our simulations. In \autoref{sec:results} when we quote summary properties for quantities that are measured on the fly in the simulations we are referring to average values for those quantities over all times $t>10\tsh$ and will quote errors as standard deviations of these quantities over the same time range.

\subsection{Diffusive Velocity}
\label{subsec:vdiff_measure}

The diffusive velocity, $\vdiff$, as it is used in \autoref{eq:edot_generic}, is meant to represent the velocity at which gas is moving into a cooling interface on small scales. We can measure this velocity in our simulations by selecting gas near the peak cooling temperature, $\Tpk$, and looking at the velocity field on either side of the interface. The diffusive velocity being the velocity at which gas is moving into the interface from the hot to the cold sides we can heuristically write this as
\begin{equation}
    \label{eq:vdiff_heuristic}
    \vdiff \approx v_{\rm in, hot} - v_{\rm in, cold}\, ,
\end{equation}
where $v_{\rm in, hot}$ is the velocity on the hot side pointing in the direction perpendicular to the interface and $v_{\rm in, cold}$ is the same on the cold side.

To formalize this heuristic we define the direction into the cooling layer by the direction pointing down the temperature gradient, evaluated at $T=\Tpk$
\begin{equation}
    \nhat \equiv - \frac{\mathbf\nabla T}{|\nabla T|}
\end{equation}
and define the velocity into the layer as 
\begin{equation}
    \label{eq:vin_local_def}
    \vinloc \equiv \vvec\cdot\nhat \, .
\end{equation}
In order to use the above in a meaning akin to \autoref{eq:vdiff_heuristic} we must look at its difference from one side of the layer to the other. To this end we take its gradient in the direction perpendicular to the layer and multiply by some length scale, $\lambda$, which should be representative of the thickness of the layer
\begin{equation}
    \label{eq:vdiff_measure}
    \vdiff = -\lambda\, \nhat\cdot\nabla \vinloc \,.
\end{equation}
The minus sign originates from our wanting $v_{\rm in,hot}-v_{\rm in,cold}$ and not vice-versa.

In practice in our simulations we compute \autoref{eq:vdiff_measure} for all gas with $1/\fwindow<T/\Tpk<\fwindow$ for a `window' factor $\fwindow = 10^{0.1}\approx 1.26$ as a proxy for gas near the peak cooling temperature. For cells which meet this criterion we compute $\nhat$ locally using centered differences of the temperature field. We then use this local estimate of $\nhat$ to compute $\vinloc$ for all cells sharing faces with the target cell (we are ignoring the small changes in the definition of $\nhat$ from cell to cell). We then compute $\nabla\vinloc$ using centered differences and compute \autoref{eq:vdiff_measure} using the previously measured $\nhat$ and $\lambda = 1.2\Delta x$. We justify this choice of $\lambda$ below by showing that, along with our measurements of $\Aint$ and \autoref{eq:edot_generic}, it reproduces the observed cooling rates in the simulations.

\subsection{Surface Area}
\label{subsec:surface_measure}

To measure the surface area of the interface between the hot and cold gas we implement the marching cubes algorithm of \citet{Lewiner03_marching_cubes} as an on-the-fly calculation in \ak. This algorithm identifies iso-value surfaces of a scalar field defined on a three-dimensional cartesian grid by considering a group of 8 neighbor cells and determining whether the scalar field at each cell lies above or below the desired iso-value. Depending on these values a triangulation is assigned to this group of cells that is representative of the iso-surface in this region. Given there are 8 cells and each can lie above or below the iso-contour there are $2^8 = 256$ different cases, 14 of which are not related to one another through rotation \citep{lorensen87}. Of these 14 cases, 7 have topologically distinct tilings as identified by \citet{Chernyaev95} and accounted for efficiently in the \citet{Lewiner03_marching_cubes} algorithm.

We use the temperature field defined on the grid and $\Tpk$ as the iso-value to pick out the contour of peak cooling gas. We validate our implementation against that of \citet{scikit-image}. The area of the interface between the hot and cold gas can be considered to be a function of scale, $\Aint(\ell)$. With our above choices for the on-the-fly measurement we have implicitly chosen to consider $\Aint(\dx)$. We do this because we are also effectively measuring $\vdiff$ on the grid scale, as we detail in \autoref{subsec:vdiff_measure}.

\begin{figure*}
    \centering
    \includegraphics[width=\textwidth]{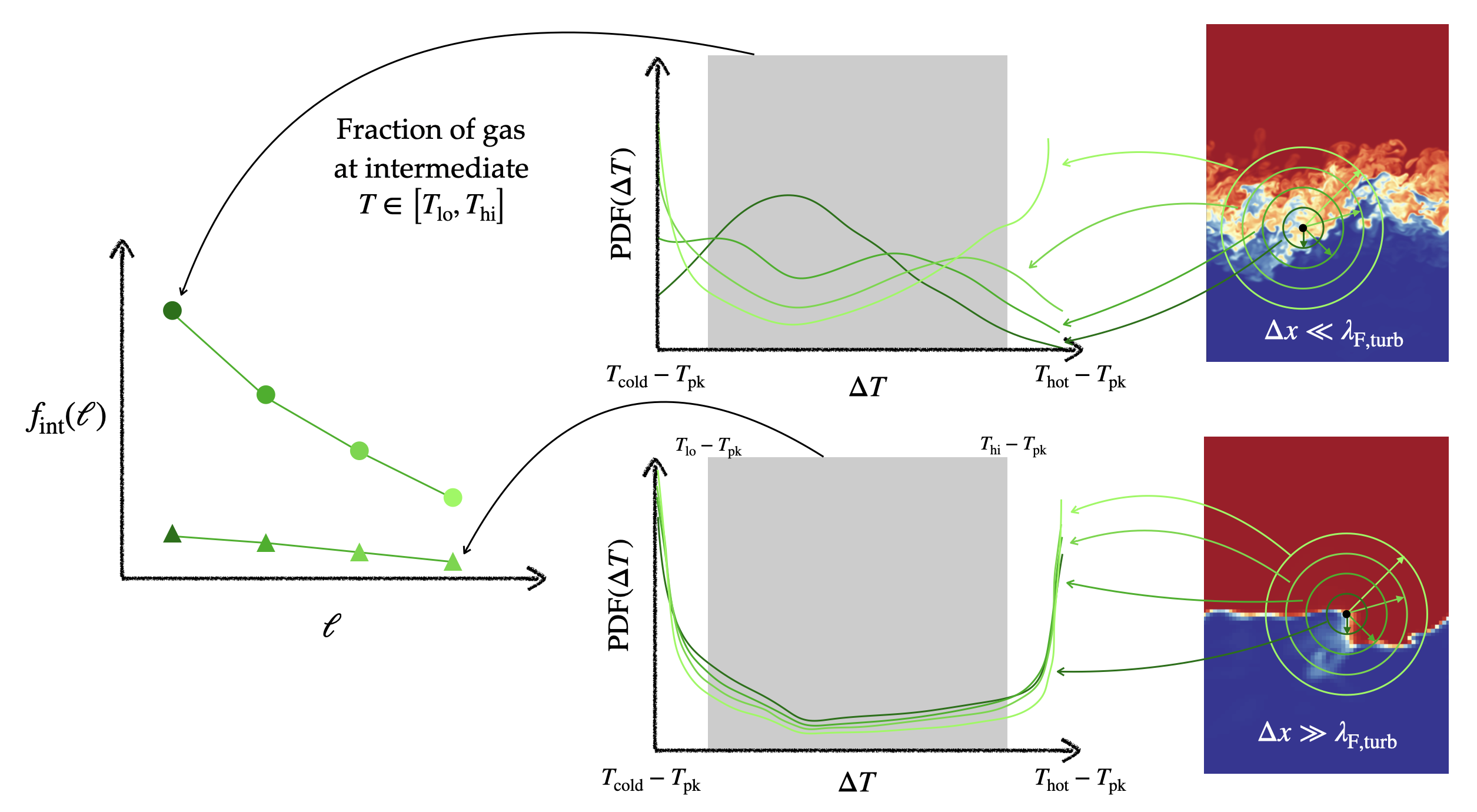}
    \caption{We provide a schematic description of our temperature ``structure function'' measurement and how it probes the phase distribution of gas near the mixing layer. \textit{Right}: We pick points near the peak temperature $\Tpk$ (black points) and compare it with the temperature of gas a distance $\ell$ away for varying $\ell$ (represented by different green circles). \textit{Middle}: We then calculate the probability distribution function (PDF) of temperature differences on a given scale. We then consider the fraction of gas at intermediate temperatures, between $T_{\rm lo}$ and $T_{\rm hi}$ (represented by the shaded region). \textit{Left}: We investigate how this fraction varies as a function of scale, $\ell$. Simulations which resolve mixing are expected to resolve scales where a significant fraction of the gas is at intermediate temperatures (top right) whereas those that do not will always have large temperature differences on all scales (bottom right).}
    \label{fig:Tsf_schematic}
\end{figure*}

\subsection{Structure Functions}
\label{subsec:sf_measure}

We now describe the measurement of the second-order velocity SFs of the mixing layers as described by \autoref{eq:vsf2_def}. Since the averaging process indicated by the angle brackets and the volume integral in this equation are both linear operations, we can calculate the second-order SFs for $v_x$, $v_y$, and $v_z$ separately as
\begin{equation}
    \label{eq:vsf_vi}
    {\rm SF}_2(v_i; \ell) = \left[\frac{1}{\Vmix}\int_{\Vmix} \left\langle|v_i(\mathbf{r}) - v_i(\mathbf{r} + \mathbf{l})|^2 \right\rangle_{|\mathbf{l}| = \ell} d\mathbf{r} \right]^{1/2}
\end{equation}
and get the SF of the overall velocity field as ${\rm SF_2(\vvec; \ell) = \sqrt{\sum_i {\rm SF}_2(v_i;\ell)^2}}$.

We calculate \autoref{eq:vsf_vi} by selecting the volume, $\Vmix$, within the simulation domain to be representative of the part of the mixing layer where turbulence is causing mixing between the phases. We define this volume by all gas that is within a certain range in $z$ that is close to peak cooling gas. Specifically, 
\begin{equation}
    \label{eq:Vmix_def}
    \Vmix \equiv \{(x,y,z) :\tilde{z}_{\rm lo}< z <\tilde{z}_{\rm hi}\}    
\end{equation}
where
\begin{equation*}
    \tilde{z}_{\rm lo} = 0.9\times(z_{\rm lo} +0.5) - 0.5 \, ,
\end{equation*}
\begin{equation*}
    \tilde{z}_{\rm hi} = 1.1\times(z_{\rm hi} + 0.5) - 0.5 \, , 
\end{equation*}
\begin{equation*}
    z_{\rm lo} \equiv \min(Z_{\rm pk}) \,\,\, , \,\,\, z_{\rm hi} = \max(Z_{\rm pk}) \, ,
\end{equation*}
and
\begin{equation*}
    Z_{\rm pk} \equiv\{z : 1/\fwindow<T(x,y,z)/\Tpk < \fwindow\}
\end{equation*}
with $\fwindow = 10^{0.1}$. In other words, we select all gas whose $z$ position is within 10\% of the $z$-position of any peak cooling gas, where the percent difference is measured with respect to the bottom edge of the domain.

Once $V$ is defined for a given snapshot we sample $10^4$ ($10^3$ for the $\nres = 512$ simulations\footnote{Each deviation vector comparison provides roughly a factor of $8$ more point-wise comparisons in each higher resolution. As the comparisons become expensive to calculate in the $\nres=512$ case, we perform fewer of them, though the results are still statistically well-converged.}) deviation vectors, $\mathbf{l}$, whose norms are distributed as $|\mathbf{l}| \sim \ell^{-1/2}$ up to a scale $\Lbox$ and are otherwise randomly oriented in space. This distribution is chosen so that we have a significant number of data points at small separations in order to accurately measure the SF on these scales. For each such deviation we offset the whole grid by this vector (accounting for periodicity in $x$ and $y$) and calculate differences over all pairs of points in the grid which both lie within the volume $\Vmix$ (as deviations in the $z$ direction can lead to comparison to points outside the volume $\Vmix$). We compute the mean over all such differences and add this value to a histogram which is binned logarithmically in spatial scale.

The result of this calculation is the ${\rm SF}_2(v_i;\ell)$ for each snapshot. As we justify further in \autoref{app:isotropy}, the $y$-component of the velocity field is the only fair tracer of the turbulence, especially on large scales. Therefore, when we refer to the turbulent SF as measured in our simulations below we will mean
\begin{equation}
    \label{eq:vsf_measure}
    \vtl \equiv \sqrt{3}\,{\rm SF}_2(v_y;\ell) \, ,
\end{equation}
where the $\sqrt{3}$ intends to account for turbulent motions in the $x$ and $z$ directions that are not easily probed by structure functions due to large-scale gradients (see \autoref{app:isotropy}).

\subsection{Temperature Structure}
\label{subsec:temp_sf}

We compute another summary quantity which is somewhat analogous to a structure function for the temperature field and is walked through schematically in \autoref{fig:Tsf_schematic}. For gas near the peak cooling temperature we would like to measure properties of the distribution of temperature differences as a function of scale, $\ell$. In particular consider the set of temperature differences:
\begin{equation}
    \label{eq:dt_set_def}
    \Delta T (\ell) \equiv 
    \left\lbrace T(\mathbf{x}+\mathbf{r}) - T(\mathbf{x}) : |r|= \ell\,\, \& \,\,T(\mathbf{x}) = \Tpk\right\rbrace 
\end{equation}
between gas at (or near) the peak cooling temperature and gas separated by a distance $\ell$.

Defining $Q_{\ell} (x)$ as the cumulative distribution function (CDF) of this set, then the fraction of gas at intermediate temperatures as
\begin{equation}
    \label{eq:fxl}
    \fint(\ell) = 
    Q_{\ell}\left(\Thi -\Tpk\right) - 
    Q_{\ell}\left(\Tlo-\Tpk\right) \, ,
\end{equation}
for some $\Tlo$ and $\Thi$. In practice, we choose $\Tlo = 1.5\times \Tcold$ and $\Thi = \Thot/1.5$ so that $\fint$ represents the fraction of gas that is more than 50\% away from one of the stable phases in temperature.

Our measurement of this function mirrors the calculation of the velocity SF except that we keep track of the whole distribution of differences and we do not compute differences over the whole volume $\Vmix$ defined above but instead, as is suggested by \autoref{eq:dt_set_def}, restrict to differences where one point is near the peak cooling temperature. In particular, we consider all points with temperature $1/\fwindow<T/\Tpk < \fwindow$ as `near' peak cooling and allow for all possible deviation vectors $\mathbf{l}$ such that $|\mathbf{l}| = \ell$.

\subsection{Summary Quantities \& History Variables}
\label{subsec:misc_measure}

We measure a few other summary quantities for all snapshots of the simulations, though as above we only present the results of these calculations for $t>10\tsh$. These calculations include:
\begin{itemize}
    \item[1.] We compute the average pressure in 100 bins in temperature spaced logarithmically between $\Tcold$ and $\Thot$.

    \item[2.] We compute the fraction of the simulation volume that is contributed by each of these temperature bins. In other words we compute the histogram $dV/d\log T$.

    \item[3.] For each constant-$z$ slice of the domain of each simulation, we compute volume averages in velocity, pressure, temperature, and density as well as covariances between the different velocity fields and the vertical turbulent ($\Rzz$) and mean ($\left\langle \rho v_z\right\rangle \left\langle v_z\right\rangle$) stress (defined after \autoref{eq:reynolds_momz}). When we average the resulting profiles across time we shift each profile to the position where the average velocity in the shear direction is zero ($\left\langle v_x \right\rangle = 0$). This is to account for the movement of the mean position of the TRML up and down with time.
\end{itemize}

\ak allows the user to output quantities that are measured on-the-fly as `history variables.' We output all history variables at $10^4$ equally spaced points in time throughout the simulation. We treat the measurement of $\vdiff$ and $\Aint(\dx)$ as history variables along with the total net cooling rate
\begin{equation}
    \edot \equiv \int_V \left(\edotc - \edoth \right) dV \, .
\end{equation}

\begin{figure}
    \centering
    \includegraphics{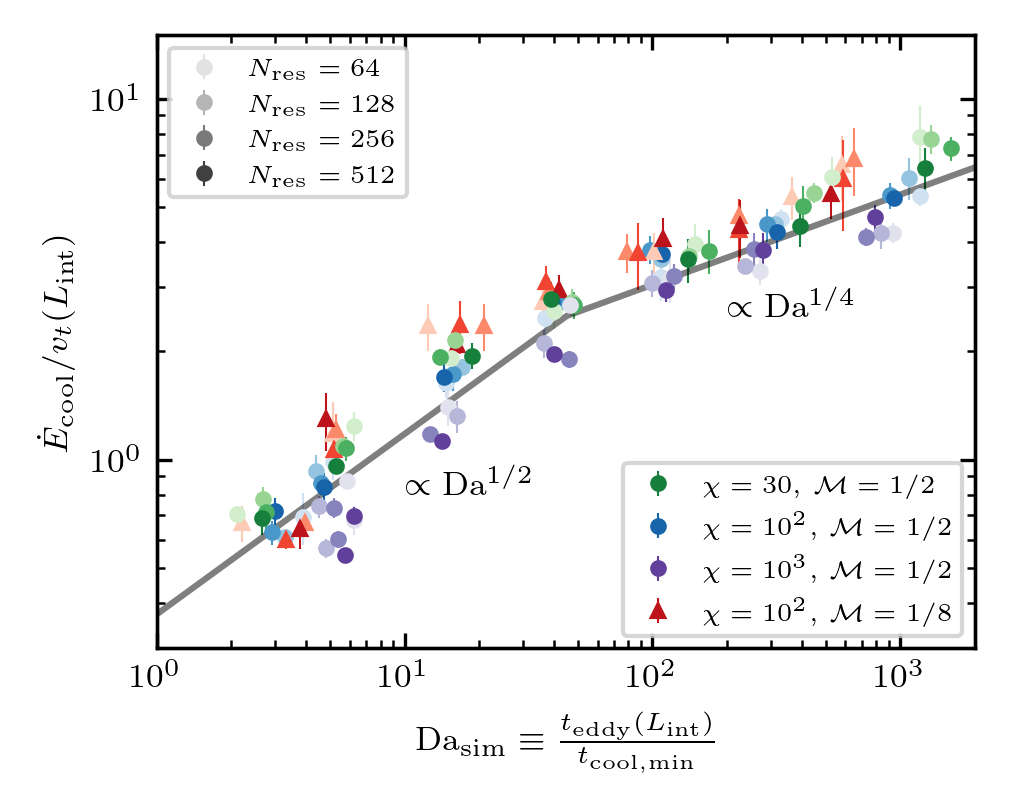}
    \caption{The total cooling rate, compensated by the turbulent velocity at the integral scale, as it varies with $\Da$ for our simulation suite. More opaque points are higher resolution simulations. \textbf{Simulations with varying $\chi$ and $\mach$ are indicated in the legend at the bottom-right}. Thin gray lines are shown to guide the eye for $\edot \propto \Da^{1/2}$ and $\edot \propto \Da^{1/4}$ scalings with a transition around $\Da \approx 50$.}
    \label{fig:edot_xi}
\end{figure}

\section{Results}
\label{sec:results}
We now present our main results. In \autoref{subsec:res0} we confirm the resolution independence of total cooling observed in past work, while also noting two distinct scaling regimes with $\Da$, which is investigated in \paperii. \autoref{subsec:cooling_and_hflux} demonstrates that this resolution independence is enabled by a cancellation between numerical diffusivity (controlling $\vdiff$) and numerical viscosity (controlling $\Aint$). However, \autoref{subsec:resolving_lcool} and \autoref{subsec:phase} demonstrate that this cancellation does not extend to the phase structure: accurate representation of the temperature distribution requires resolving $\lcool$. We discuss the physical origins of the phase structure in \autoref{subsec:phase_origin}.

\subsection{Resolution Independence of Total Cooling}
\label{subsec:res0}

We begin by examining the total cooling rate, $\edot$, the primary integrated diagnostic of mixing layer energetics. \autoref{fig:edot_xi} shows $\edot$ as a function of $\Da$, as measured in the simulations. This measurement is done using the turbulent SF measurement described in \autoref{subsec:sf_measure} to get the eddy turnover-time on a scale $\Lint$ (which we define as the scale which maximizes $\vtl$):
\begin{equation}
    \label{eq:Da_measure}
    \Dasim = \frac{\teddy(\Lint)}{\tcoolmin} \, .
\end{equation}
This will be used to indicate the value as measured in the simulations throughout the rest of the work. In \paperii, we discuss other methods of measuring $\Dasim$, its relationship to $\xi$, and the use of $\tcoolmin$ as the relevant cooling time.

Although all of our simulations lie in the $\Da \geq 1$ regime, we see in \autoref{fig:edot_xi} that the two distinct regions of scaling behavior reported in previous studies still appear \citep{FieldingFractal20,Tan21}. At moderate values of $\Da$ we see $\edot\propto \Da^{1/2}$ which transitions to $\propto \Da^{1/4}$ only at $\Da\gtrsim50$. In \paperii, we show that the origin of this change is a suppression in $\Aint$ due to the ram-pressure of the inflowing hot gas.

More importantly for the current work, it is also clear that $\edot$ is nearly independent of resolution across the different simulations. This feature is also observed in past works \citep{FieldingFractal20,Tan21} but is somewhat strange given that we are fundamentally studying a diffusive process and there is no explicit diffusion, only numerical diffusion. We return to this below.

\begin{figure}
    \centering
    \includegraphics{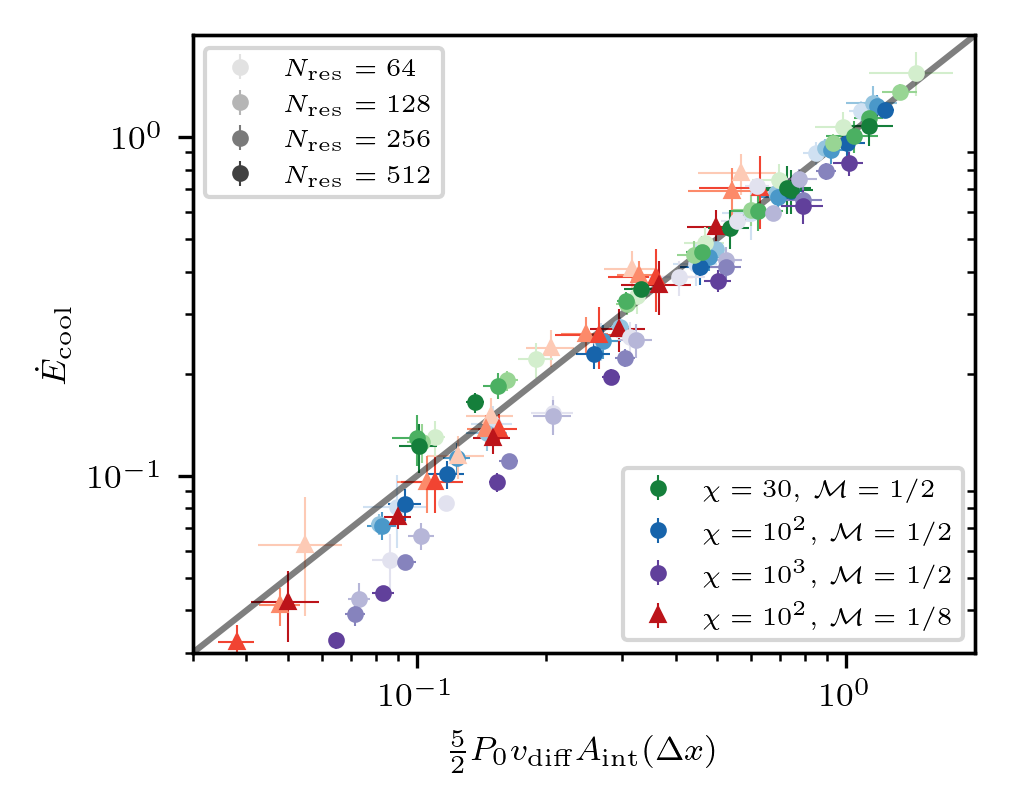}
    \caption{We show the total cooling rate versus the enthalpy flux into the layer as given by \autoref{eq:edot_generic} with $\vdiff$ and $\Aint$ measured in the simulations as detailed in \autoref{sec:measurement}. Each point corresponds to a separate simulation with the same style as \autoref{fig:edot_xi}. The gray line is one-to-one.}
    \label{fig:edot_vs_hflux}
\end{figure}

\begin{figure*}
    \centering
    \includegraphics[width=\textwidth]{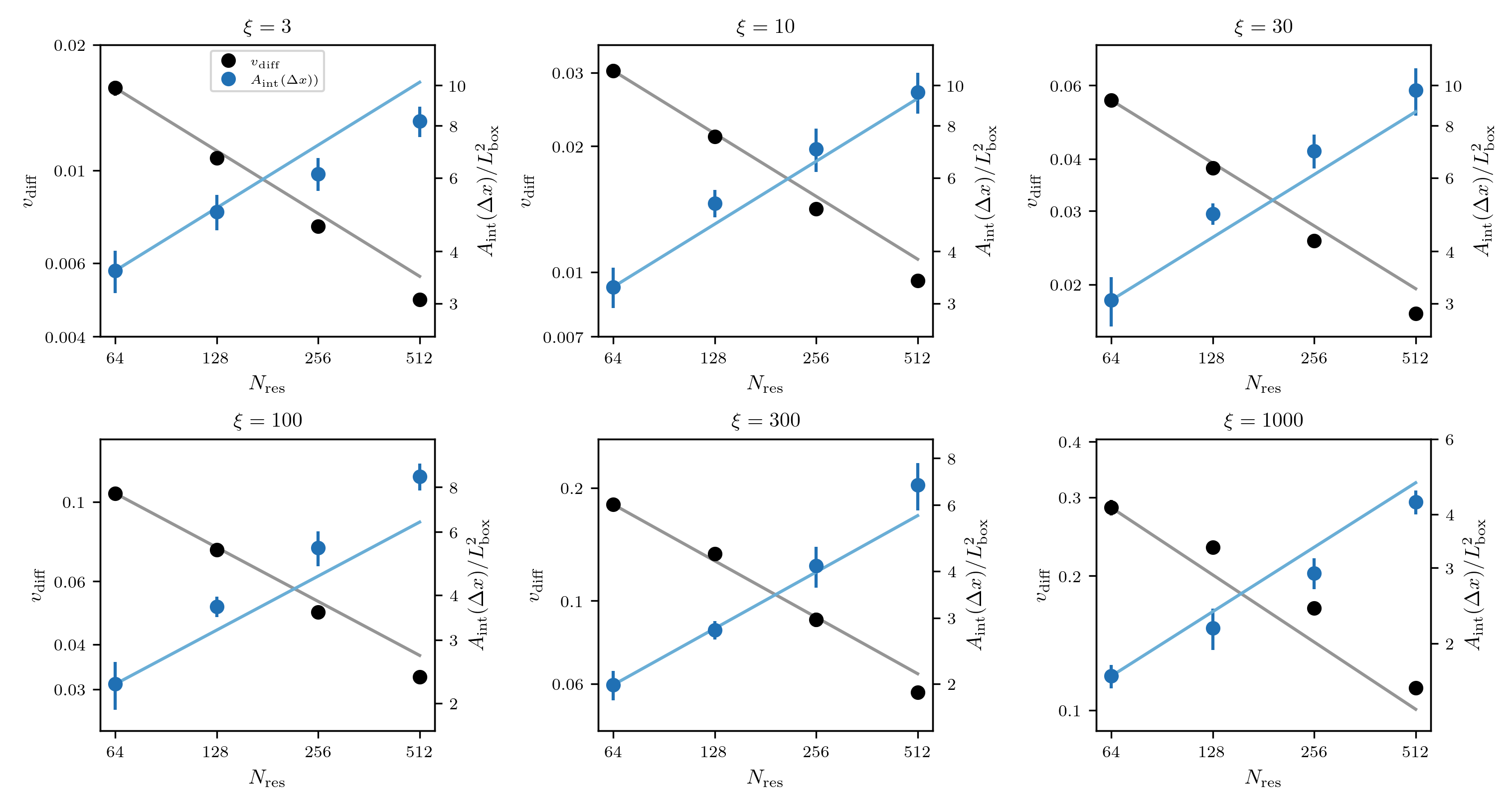}
    \caption{We show the behavior $\vdiff$ and $\Aint$ as a function of resolution ($\nres$) in the $\mach = 1/2$ simulations. We show different $\xi$-valued simulations in each panel, with labels above each panel. $\vdiff$ is shown in the black points with corresponding $y$-axis on the left, and $\Aint(\dx)$ is shown as blue points with corresponding $y$-axis on the right of the panel with each panel $y$-axis covering the same logarithmic range in each panel. Gray (blue) lines are showing scaling relations $\propto \nres^{-1/2}$ ($\propto\nres^{1/2}$).}
    \label{fig:scaling_cancel}
\end{figure*}

\subsection{The Cancellation: Numerical Diffusivity vs. Numerical Viscosity}
\label{subsec:cooling_and_hflux}

In \autoref{sec:theory} we discussed how the total cooling in the layer must be balanced by the flux of high specific-entropy gas into the layer, as written in \autoref{eq:edot_generic}. We measure the components of this flux, specifically $\vdiff$ and $\Aint$ in our simulations as described in \autoref{sec:measurement}. The results of these calculations are given in \autoref{fig:edot_vs_hflux}, where we show $5 \Pbar \vdiff \Aint(\dx)/2$ versus the measured total cooling rate in the simulations.

The agreement is excellent across all simulations, validating both the theoretical framework of \autoref{eq:edot_generic} and our measurement techniques. The measurement of $\vdiff$ includes a calibrated O(1) prefactor (the length scale $\lambda = 1.2 \Delta x$ in \autoref{eq:vdiff_measure}), but crucially, the same prefactor works across simulations spanning three orders of magnitude in $\tcool$, a factor of 8 in resolution, and a factor of $\sim 30$ in over-density. The energy budget of turbulent radiative mixing layers is, therefore, definitively controlled by the diffusive enthalpy flux.

Having established that $\edot = \tfrac{\gamma}{\gamma -1}P \vdiff \Aint$, we can now ask why $\edot$ is resolution-independent even though both factors depend on resolution. \citet{Lancaster24a} proposed that $\vdiff$ and $\Aint$ scale oppositely with resolution: numerical diffusion decreases at higher resolution (lowering $\vdiff$), but reduced numerical viscosity allows more small-scale structure (raising $\Aint$). If these effects cancel, $\edot$ would be resolution-independent. \autoref{fig:scaling_cancel} tests this hypothesis directly, showing $\vdiff$ and $\Aint(\Delta x)$ as functions of resolution for the $\mach=1/2$ simulations. The cancellation is remarkably precise: $\vdiff \propto \nres^{-1/2} \propto \dx^{1/2}$ while $\Aint(\Delta x) \propto \nres^{1/2} \propto \dx^{-1/2}$, so their product is resolution-independent. \emph{This is the root of our puzzle:} total cooling is converged not necessarily because the physics is resolved, but because two numerical artifacts---numerical diffusivity and numerical viscosity---cancel in their effects on $\edot$. This scaling continues even when we resolve $\lcool$, which we return to in \autoref{subsec:res_req}. As we show in \autoref{app:res_dep}, this convenient cancellation does not persist over all regimes of parameter space explored by our simulations.

The area scaling is explained heuristically by a fractal model where
\begin{equation}
    \label{eq:aint_frac}
    \Aint(\ell) \approx \Lbox^2 \left(\frac{\ell}{\Lbox}\right)^{-d}
\end{equation}
with $d = 1/2$ the excess fractal dimension. The origin of this value for $d$ is as yet unclear, but this same fractal scaling was also observed in past work \citep{FieldingFractal20,Lancaster21b,Lancaster24a}. We investigate behavior of $\Aint(\ell)$ on all measured scales in \paperii.

\begin{figure}
    \centering
    \includegraphics{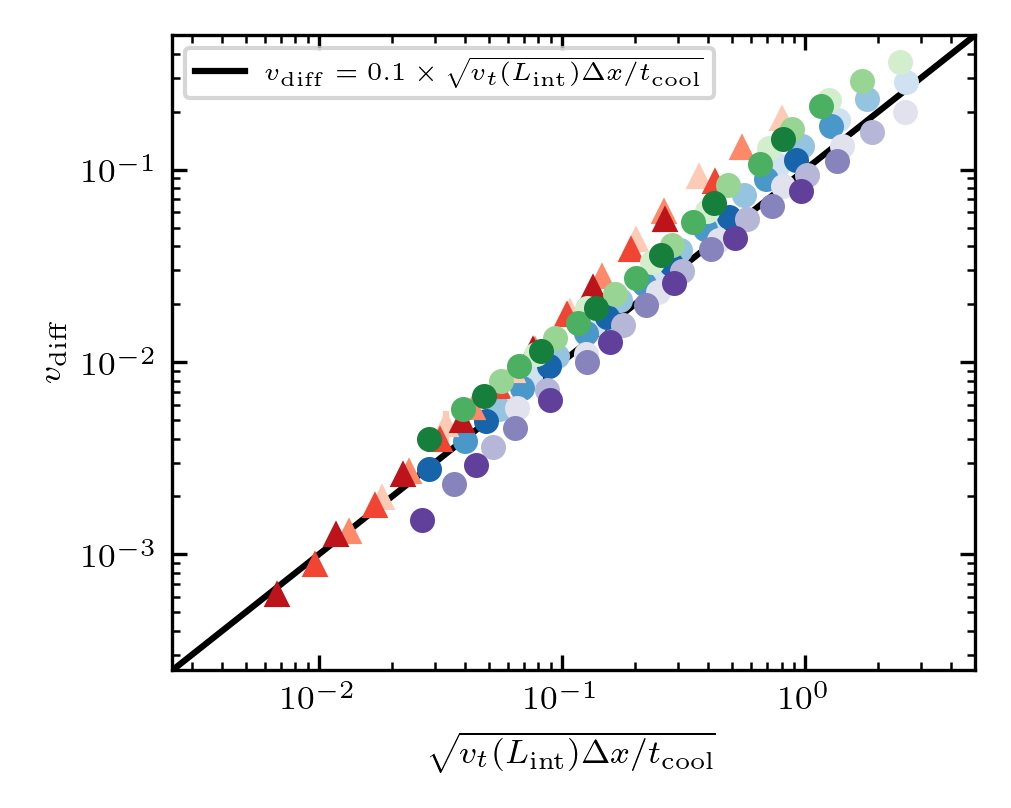}
    \caption{The diffusive velocity, $\vdiff$, versus the value expected for the diffusive velocity if it is determined by numerical diffusivity with velocity scale given by the order of velocity fluctuations on the grid $\vtL = \sqrt{\sum_i {\rm SF}_2(v_i)}$ and length scale $\dx$. The points are for all simulations in our suite and include error bars on $\vdiff$, which are generally smaller than the points.  The black line is a linear relation with proportionality constant of 0.1, as indicated by the legend. Point styles are the same as in \autoref{fig:edot_xi}.}
    \label{fig:vdiff_numeric}
\end{figure}

We might expect that since there is no explicit diffusion in our simulations $\vdiff$ should be given by an equation of the form of \autoref{eq:vdiff_scale} with the thermal diffusivity, $\alpha_T$ replaced with an appropriate numerical diffusivity. It is difficult to derive an exact formula for the numerical diffusivity of a Godunov scheme for the full Euler Equations\footnote{Though empirical calculations of effective dissipative terms for whole simulation domains have been performed \citep[e.g.][]{Grete23_num_disp}}. However, in our nearly isobaric simulations the specific entropy acts approximately as a passive scalar. For passive scalar advection, the numerical diffusivity of a Godunov scheme scales as $\anum \sim v\dx$, where $v$ is the advection velocity \citep{ToroRiemann,NumericalRecipes}. If we extend this analogy to the specific entropy we would expect the numerical diffusivity to be given by
\begin{equation}
    \label{eq:alpha_num}
    \anum \approx v_t (L)\dx \, .
\end{equation}
That is, we choose the velocity scale as the magnitude of velocity fluctuations in the ``lab'' frame of the grid (in analogy to the advection velocity $v$ in the case of the advection equation), and the length scale as the resolution scale.

If this were the numerical diffusivity, following \autoref{eq:vdiff_scale} we would expect the diffusive velocity to be given by
\begin{equation}
    \label{eq:vdiff_num_pred}
    \vdiffn \approx \sqrt{\frac{v_t(\Lint)\dx}{\tcoolmin}} \, .
\end{equation}

\autoref{fig:vdiff_numeric} confirms this prediction: the measured $\vdiff$ is proportional to \autoref{eq:vdiff_num_pred} with proportionality constant 0.1. The linear relationship, holding over two orders of magnitude in the predicted value, demonstrates that diffusion in these simulations is fundamentally numerical. The O(1) prefactor reflects geometric uncertainties in defining $\anum$; the key point is that a \textit{single prefactor works across all simulations}! This confirms that diffusivity in these simulations is entirely numerical, with a consistent functional form across all parameters studied. The resolution independence of $\edot$ is thus not necessarily a sign that the physics is resolved—it is possibly a coincidence arising from the cancellation of two resolution-dependent quantities. Understanding whether this cancellation is physical or a coincidence of numerics requires either (i) a demonstration that this does not hold in similar set ups with different numerical methods (e.g. moving mesh) or (ii) an understanding of the origin of the fractal dimension of the interface, $d$. A first principles explanation for the value of $d$ is elusive even in the problem of scalar mixing (no cooling) where values ranging from $1/3-2/3$ are quoted in the literature for different regimes \citep[e.g.][]{CPS91,SRM89,MoninYaglomBook}. We now ask: what is required for genuine resolution of turbulent mixing?

\begin{figure}
    \centering
    \includegraphics{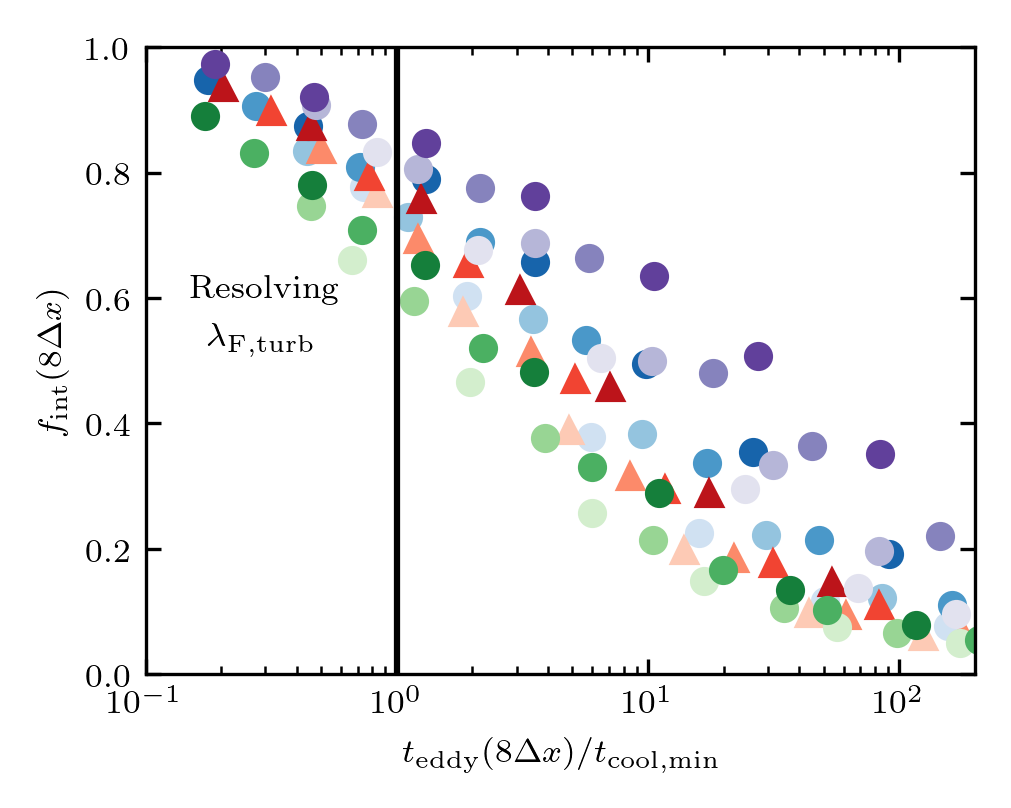}
    \caption{The fraction of gas at intermediate temperatures, $1.5\times \Tcold < T < \Thot/1.5$, separated by a distance $8\dx$ from gas at $\Tpk$ as a function of $\teddy(8\dx)/\tcoolmin$. $8\dx$ is chosen as the minimum scale considered to be well-resolved (see \paperii). The vertical black line corresponds to simulations which resolve $\lcool$, at which point $\fint\gtrsim 80\%$. Point styles are the same as in \autoref{fig:edot_xi}.}
    \label{fig:teddy_Tsf}
\end{figure}

\subsection{The Turbulent Field Length $\lcool$ as a Resolution Criterion}
\label{subsec:resolving_lcool}

In \autoref{sec:theory} and \autoref{fig:resolution_schematic} we argued that once one is able to resolve $\lcool$, turbulent diffusion is able to compete with cooling on a realistic timescale. Once this scale is resolved, we would then expect to see the hallmark of that diffusion: a smooth transition in the phase distribution of the gas. We note again that in these simulations the actual smoothing of the interface on the smallest scales is mediated by numerical diffusion, which plays the role that physical diffusion would in reality. The hallmark of the layer being resolved in the sense we mean here is that the \textit{effective} turbulent diffusivity is realistically represented.

We use the temperature ``structure function'' measurement described in \autoref{subsec:sf_measure} in order to quantify this transition. In particular, in \autoref{fig:teddy_Tsf} we show the fraction, $\fint$, of gas that is separated from gas at $\Tpk$ by a distance of $\ell\approx 8\dx$ and is at intermediate temperatures $T \in [1.5\times\Tcold, \Thot/1.5]$ (see \autoref{eq:fxl} and \autoref{fig:Tsf_schematic}). We choose $\ell =8\dx$ as we consider this to be the minimal scale still in the ``well-resolved'' regime of the turbulence (on scales smaller than this, turbulent motions are damped by numerical viscosity, see \paperii).

If the gas is truly multiphase on this minimally resolved scale, then a large fraction of the gas on this scale will be near $\Thot$ or $\Tcold$ and $\fint$ will be small. If the gas is efficiently mixed by turbulent diffusion, then most of the gas will be at intermediate temperatures, and $\fint$ will be large. We see in \autoref{fig:teddy_Tsf} that when a simulation resolves $\lcool$ (left of the vertical black line) nearly all ($\gtrsim 80\%$) of the gas on this scale is at intermediate temperatures. This trend is also visually apparent in the slices of \autoref{fig:multi_panel} where we indicate how well-resolved $\lcool$ is in the bottom-right of each panel.

It is also clear from \autoref{fig:teddy_Tsf} that the transition from $\fint\approx 0$ to $\fint\approx1$ is not an instantaneous transition once $\lcool$ is resolved. This is to be expected. As we described in \autoref{sec:theory}, there are two effects competing to change the entropy of a gas parcel in the TRML: mixing and cooling. When $\lcool$ is not well resolved, the time-scales over which mixing acts are not faithfully represented in the simulation, the competition between these two effects is therefore not well-resolved, and the transition between the two phases is sharp ($\fint \approx 0$). As we move to higher resolution and $\lcool$ becomes better resolved, the competition between mixing and cooling becomes increasingly fair. This is manifested in the fact that $\fint \approx 40\%$ even when $\teddy \approx10\,\tcoolmin$ in \autoref{fig:teddy_Tsf}. The smooth progression of this transition should therefore not be considered as evidence against its fundamental importance.

In summary, simulations which resolve $\lcool$ show a smoothing out of the local phase structure near the interface. This is excellent evidence that $\lcool$ is a meaningful scale on which to consider the turbulence as efficiently mixing the phases. We next investigate the resolution dependence of the phase-structure of intermediate temperature gas in the mixing layer and how this compares to the resolution of $\lcool$.

\begin{figure*}
    \centering
    \includegraphics[width=\textwidth]{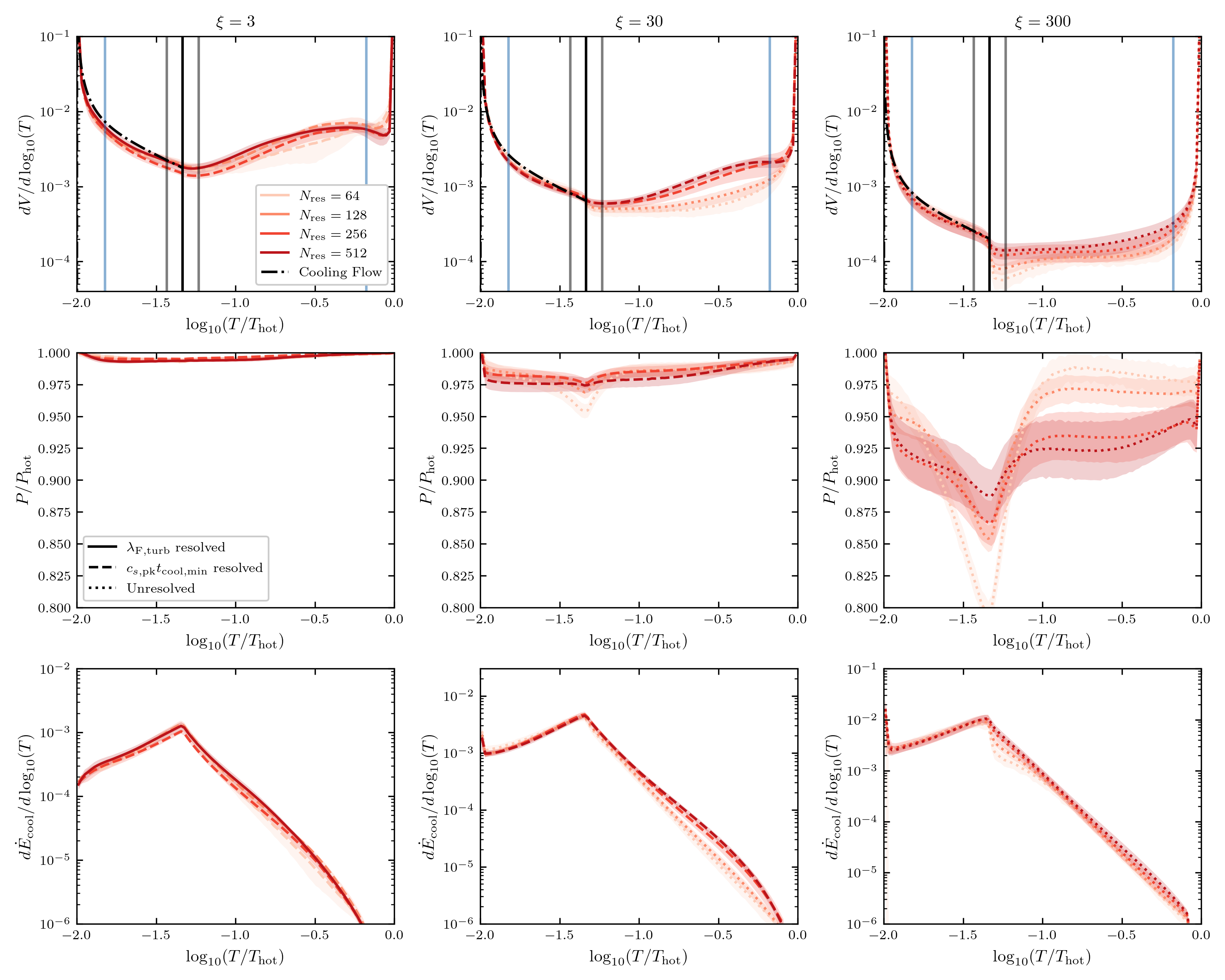}
    \caption{The volume weighted temperature PDF (top row) the average pressure as a function of temperature (middle row), and the cooling-weighted temperature PDF (bottom row) in the $\xi = 3,\, 30,\, \&\, 300$ simulations (left, middle, and right columns respectively) at $\mach = 1/8$. Higher resolution simulations are represented by darker lines, as given by the legend in the top left panel. Simulations with $\lcool > 8\,\dx$ are shown with solid lines, simulations with $\cspk \tcoolmin > 8\, \dx$ are shown as dashed lines, all others are shown as dotted lines. Black dot-dahsed lines on the left-hand side of the top panels represent the expected phase distribution for a pure cooling flow, see text for more details. In the top panels vertical lines indicate $\Tpk$ (black), range of temperatures considered to be ``near'' $\Tpk$ (gray) and ``intermediate'' temperature range used in \autoref{fig:teddy_Tsf} (blue).}
    \label{fig:phase_plot}
\end{figure*}

\subsection{Convergence of Phase Structure}
\label{subsec:phase}

In order to consider our simulations as well resolved we would like them to not only be converged in integrated properties, like $\edot$, but also in aspects of the distribution of the gas, such as its phase structure, and therefore its emissivity as a function of temperature. As has been discussed in several past works \citep{McCourt18,FieldingFractal20,TanOh21,Abruzzo24}, one way in which the phase distribution could be inaccurately represented is if the simulation does not resolve $\cspk\tcoolmin$, the scale that a sound wave crosses in a cooling time. If this scale is unresolved then the gas in a given cell loses pressure support to cooling more quickly than it can be compressed and re-pressurized by the surrounding gas, resulting in artificial dips in the pressure.

In \autoref{fig:phase_plot} we show the volume-weighted temperature distribution, the average pressure as a function of temperature, and the cooling-weighted temperature distribution (measurement details in \autoref{subsec:misc_measure}) in the $\xi= 3,\, 30,\, \&\, 300$ simulations with $\mach = 1/8$. Simulations that resolve  $\lcool$ ($\cspk\tcoolmin$) by more than 8 cells are shown as solid (dashed) lines and all others are shown as dotted lines. As the turbulence in these simulations is always subsonic in the mixed gas, $v_t < \cspk$, resolving $\lcool$ is a more stringent constraint than resolving $\cspk\tcoolmin$ in subsonic mixing layer simulations. We choose the $\mach = 1/8$ simulations here as there is more separation in scale between $\cspk\tcoolmin$ and $\lcool$ in these runs. Some points of note:
\begin{itemize}
    \item[1.] \emph{Physical pressure dips exist at high resolution}. \citet{FieldingFractal20} demonstrated that artificial dips in the gas pressure at intermediate temperature exist when $\cspk\tcoolmin$ is not well resolved. However, there is a well resolved dip in the thermal pressure of the gas at intermediate temperatures, even when $\cspk\tcoolmin$ is well resolved (see central panel, darkest dashed line). This is not a resolution artifact but a physical consequence of turbulent pressure support \citep{Ji19,Abruzzo24,Sharma25}. This is in contrast with 1D models which do not include turbulent pressure support \citep{CFB23}.
    
    \item[2.] \emph{Under-resolving $\cspk\tcoolmin$ distorts $T>\Tpk$ gas.} When $\cspk\tcoolmin$ is unresolved (dotted lines), gas at $T>\Tpk$ is under-represented in the volume distribution and preferentially over-pressurized (see particularly top-right and middle-right panels). While these effects tend to balance one another in terms of the total emission from the layer (bottom panels), the simulations that don't resolve $\cspk\tcoolmin$ are clearly unresolved in the temperature structure of emission from the layer.
    
    \item[3.] \emph{$T<\Tpk$ gas is more robust.} In the same unresolved simulations, at $T<\Tpk$ the gas volume distribution is well-resolved, but the gas pressure is still slightly above its converged value (though less than in the case of the hotter gas).

    \item[4.] \emph{Resolving $\lcool$ refines the emission distribution.} Resolving $\lcool$ results in small but significant ($\sim 50\%$) changes in the cooling-weighted distribution when $\cspk\tcoolmin$ is already resolved (bottom-left panel).
\end{itemize}
Overall, this implies that while low-resolution simulations may accurately predict the global energy dynamics, $\edot$ (although see Appendix~\ref{app:res_dep}), they will fail to predict the emission or absorption spectrum because the intermediate phase volume is artificially under-represented.

\subsection{Physical Origin of Phase Structure}
\label{subsec:phase_origin}

The resolved pressure dips observed in \autoref{fig:phase_plot} can be explained by the picture of \citet{Sharma25} (see also \citet{Ji19} discussion around Eqs. 19-20) in which a dip in the pressure at intermediate temperatures is supported by turbulent motions in the TRML. In particular, assuming a steady state and averaging the $z$-component of the momentum equation over the periodic dimensions and time of our simulations one finds (\citet{Sharma25} Equation 5\footnote{\citet{Ji19} Eq. 19 ignores the last term in \autoref{eq:reynolds_momz}, while the formalism of \citet{CFB23} ignores the Reynolds Stress term.})
\begin{equation}
    \label{eq:reynolds_momz}
    \frac{d}{dz} \left( \meanP + 
    \Rzz + \left\langle \rho v_z \right\rangle \left\langle v_z \right\rangle \right) = 0
\end{equation}
where $\left\langle \phi \right\rangle = \int \phi(\mathbf{x},t)\,\, dx\, dy\,dt/(L_x L_y t)$ is the average of a quantity over the horizontal extent of the layer, and time, and $\Rzz \equiv \left\langle \delta v_z \delta \left(\rho v_z \right) \right\rangle$ is the vertical turbulent stress ($\delta \phi\equiv \phi - \left\langle \phi \right\rangle$). We have confirmed that \autoref{eq:reynolds_momz} and in particular the picture of \citet{Sharma25} in which a dip in $\meanP$ is supported by a rise in $\Rzz$, holds in all of our simulations.

This analysis offers a frame in which to interrogate the over-pressurization of gas at intermediate temperatures when the simulation is not well-resolved. In particular, we find that in low resolution simulations $\Rzz$ is below the value it achieves at higher resolution (see \autoref{app:turb_support}), indicating that it is a lack of resolving turbulent support that forces the thermal pressure to higher values in order to satisfy \autoref{eq:reynolds_momz}. This is, to some degree, apparent in the bottom panels of \autoref{fig:multi_panel}, where the low resolution simulations have little indication of strong turbulent motions. This overall picture also explains why pressure dips in the fast cooling (high $\xi$) simulations are larger: their greater cooling and therefore greater bulk inflows results in greater turbulent support in the layer (larger $\vtL$ and $\Rzz$, though this effect is mild).

The final mystery of \autoref{fig:phase_plot} is the apparent resolution independence of the phase distribution of gas at $T<\Tpk$. If we suppose that the amount of gas at each temperature in this range is simply determined by the rate at which gas is being created at $\Tpk$ and cooling (i.e. mixing is not important for the change in temperature of this gas) then the temperature distribution of this gas would be the same as a one-dimensional cooling flow with a constant mass flux, $j_x = \rho v$. If we assume this flow as isobaric we can solve for the temperature of this gas as a function of radius as \citep{KK13,TanOh21}\footnote{Formally, such a flow cannot exist in steady-state in cartesian coordinates globally \citep{Dutta22}. We are circumventing this by ignoring the momentum equation, and supposing that we are simply being provided with gas at $\Tpk$ by another process.}
\begin{equation}
    \label{eq:cooling_flow}
    \frac{j_x \gamma}{\gamma - 1} \frac{dT}{dx} = -\edotc(P_0,T) \, .
\end{equation}
$dV/d\log T \propto T/(dT/dx)$ so that we may infer the phase space distribution of the gas directly from the cooling function using \autoref{eq:cooling_flow}. This distribution is shown as a black dot-dashed line in the top panels of \autoref{fig:phase_plot}.

The fact that this distribution agrees well with nearly all simulations shows us that the phase distribution of this gas is likely determined by the cooling function alone and that mixing is not as important for populating gas at these temperatures. This is in keeping with the near resolution independence of this part of temperature space since the rate at which gas is being populated at $\Tpk$ is basically determined by the global energy dynamics, which is relatively independent of resolution (see \autoref{fig:edot_xi}). This may not remain true if explicit heat conduction is included \citep[e.g.][]{TanOh21}.

\section{Discussion \&  Conclusion}
\label{sec:discussion}

We first provide a thorough summary of the resolution requirements for accurately representing the phase structure (and hence observable properties of emission from/absorption in the mixing layer) in \autoref{subsec:res_req}. In \autoref{subsec:prandtl} we discuss an interpretation of the resolution independence of our simulations (in terms of ${\rm Pr}$) and its relationship to other works in the literature which have included additional physics. In \autoref{subsec:summary} we summarize our main results.

\subsection{Resolution Requirements for Different Questions}
\label{subsec:res_req}

Even if the total dissipation (that leads to the total cooling) is correct in the simulations, it can still be true that the numerical dissipation is not an accurate representation of the turbulent dissipation that would exist in reality. As we discussed in \autoref{sec:theory}, in reactive flows like TRMLs where the reaction acts to return mixed gas to either of the two thermally stable phases, the thickness of the interface between the phases is given by the scale on which diffusion and this reaction balance, which can be quite small compared to the scales of the problem ($\lcool$ in the case of turbulent dissipation). When this scale is not resolved, the mixing that is occurring in the simulation is simply due to the advection of the interface over the grid. Even if the simulation gives the right amount of total mixing across the surface, the phase structure around the surface cannot be expected to be correct. In \autoref{subsec:resolving_lcool} and \autoref{fig:teddy_Tsf} we show this to indeed be the case.

What, then, is required to correctly resolve the phase structure of the layer? To get the phase structure right we must correctly resolve the processes which alter or maintain the phase (temperature, entropy, pressure, etc.) of a parcel of gas. In the simulations presented here these processes are (i) mixing, (ii) cooling, (iii) compression\footnote{As well as rarefaction, in the case of temperature and pressure.} by sound waves, and (iv) turbulent pressure support, which acts to prevent compression. Which of these processes do we accurately represent?
\begin{itemize}
    \item[-] \textit{Mixing}: The above discussion around resolution independence shows us that we do (on average) get the total amount of mixing correct. However, as discussed in \autoref{sec:theory} and demonstrated in \autoref{subsec:resolving_lcool} and \autoref{subsec:phase} we do not accurately represent the time-dependent action of turbulence to move a parcel of gas through intermediate phases when we do not resolve the eddies on scales $\ell \lesssim \lcool$ which act to diffuse on time-scales comparable to $\tcoolmin$.

    \item[-] \textit{Cooling}: If we consider these other processes as separate, then the resolution of the `cooling' process for an individual gas element is essentially enforced by the numerical techniques of our simulation (time-step limiting based on the reaction rate).

    \item[-] \textit{Compression}: When the minimum scale that a sound wave crosses in a cooling time  ($\cspk \tcoolmin$) is not resolved, a gas parcel is able to artificially cool to lower entropy/pressure before the hydrodynamics in the simulation is able to re-pressurize it. As has been discussed in \citet{FieldingFractal20} this leads to an unphysical drop in the pressure of gas in the mixing layer.

    \item[-] \textit{Turbulent Pressure Support}: \citet{Sharma25} demonstrated that vertical turbulent stress acts to dynamically support the gas in the mixing layer leading to a physical dip in the thermal pressure in the layer. In \autoref{subsec:phase} and \autoref{app:turb_support} we demonstrate that the lack of resolution of this turbulent support can lead to unphysical, high pressures at intermediate temperatures.
\end{itemize}

In order for the phase structure to be resolved, it seems we must resolve the turbulent mixing ($\lcool$), thermal pressure support ($\cspk \tcoolmin$), and the turbulent pressure support. We demonstrate in \autoref{app:turb_support} that the final requirement is satisfied well before the former ones for the simulations presented here\footnote{In general, we might expect this to be the case, as it is the largest turbulent motions which provide the most dynamical support, which will be resolved before $\lcool$ in $\Da > 1$ simulations.}. In \autoref{subsec:phase} we empirically show that resolving $\cspk\tcoolmin$ results in a reasonably well-resolved phase structure in the mixing layer (see \autoref{fig:phase_plot}) while the additional resolution of $\lcool$ can results in a $\sim 50\%$ change in emission properties. Therefore, we emphasize that accurate representations of observable properties of mixing layers require the resolution of $\lcool$ (or at \textit{least} $\cspk\tcoolmin$).

In \autoref{fig:scaling_cancel} we see that the resolution scaling of $\vdiff$ and $\Aint(\dx)$ continues even when $\lcool$ is well resolved (highest resolution points in the top-left, $\xi = 3$ panel of \autoref{fig:scaling_cancel}). We have shown that $\lcool$ is the scale on which the phase distribution becomes smooth and therefore that this is likely a meaningful scale on which ``effective'' turbulent diffusion can be considered as well represented. However, since numerics fundamentally controls both $\Aint(\dx)$ and $\vdiff$, resolution dependence is expected unless explicit dissipation is included and smooths out the interface at some scale \citep{TMG25}.

For the set of physics considered in these simulations (pure hydrodynamics, with a local cooling function) we conclude that resolving $\lcool$ is necessary to consider the emission properties as well resolved. Resolution of $\cspk\tcoolmin$, which is generally larger than $\lcool$, gives reasonably accurate results for the parameters studied here. However, we also demonstrate in \autoref{subsec:resolving_lcool} that the resolution of $\lcool$ shows a measurable change in the smoothness of the gas phase distribution near the interface. If we were to calculate the energy levels of a given ion species near the TRML interface, and these energy levels were significantly affected by radiative trapping in the layer, then it may be necessary to resolve the scales over which the phase distribution varies smoothly in order to correctly simulate line transfer, and therefore emission from these energy levels. Therefore, to answer certain questions, it may still be necessary to resolve $\lcool$ even if it does not make a large difference here.

\subsection{Physical Interpretation: The Role of $\Pr$}
\label{subsec:prandtl}

Is the countervailing behavior of numerical heat diffusion and numerical viscosity, that leads to the resolution independence of $\edot$ (\autoref{fig:scaling_cancel}), somewhat physical? In the purely hydrodynamic simulations we present here, it could be argued that this cancellation is showing us that what matters in TRMLs is the ratio of kinematic viscosity, $\nu$, to thermal diffusivity, $\alpha$, the so-called Prandtl number, $\Pr\equiv \nu/\alpha$. If $\nu$ can be thought of as controlling the area of the interface ($\Aint$) and $\alpha$ can be thought of as controlling the thermal diffusivity across that interface ($\vdiff$), then as long as the ratio of these quantities ($\Pr$) is held constant, the resulting dissipation across the surface is the same. Given that $\Pr \approx 1$ for numerical dissipation and turbulent dissipation alike, it may not be surprising that the simulations presented here demonstrate relative resolution independence in integrated quantities like $\edot$ (\autoref{fig:edot_xi}). However, as we show in \autoref{app:res_dep}, this resolution independence can break down in certain regimes, specifically when the resolution is so low that $\Aint$ is forced to approach its lower limit, $\Lbox^2$.

One implication of this interpretation is that resolution independence should not remain if one changes ${\rm Pr}$ by changing $\nu$ ($\alpha$) in a way that affects $\Aint$ ($\vdiff$). This point of view could be described as a ``diffusion limited'' perspective, where it is the micro-physical diffusivity which limits the amount of dissipation across the interface. This stands in contrast to a ``mixing limited'' perspective, popular in the scalar mixing literature \citep{Batchelor59,Warhaft00} in which, as long as the large-scale turbulent mixing rate is fixed or well-resolved, the amount of diffusion across the layer remains the same. The present work does not distinguish between these two regimes. Indeed, an answer to this question in the context of TRMLs will depend upon a thorough ${\rm Pr}$-varied study, which we plan to carry out in future work. As we discuss further below, when carrying out such a study in an effort to tell the difference between the diffusion and mixing limited pictures, it is important to ensure that changes to $\nu$ or $\alpha$ do not cause changes to the integral scale mixing, which requires very high resolution \citep{Kritsuk07,Grete23_num_disp}. In the rest of this section we discuss works in the existing literature which shed light on the question of whether a mixing or diffusion limited picture applies in TRMLs.

TRML simulations which include dynamically important magnetic fields \citep{ZhaoBai23,Das24,BSG23,Lancaster24a} introduce an effective ``viscosity'' through magnetic tension, which suppresses the instabilities that lead to small-scale turbulent folding. However, this does not alter the fundamental role of numerical heat dissipation. The net effect is an \textit{effective} change in ${\rm Pr}$, which produces a clear resolution dependence for $\edot$ in magnetized mixing layers (see e.g. \citet{Lancaster24a} or Figure A2 left panel, red curves of \citet{ZhaoBai23}). These same simulations also show reduced turbulent velocities, with the relationship between turbulent velocity ($\vtL$) and total cooling ($\edot$) approximately preserved \citep{Das24}. This fact would normally be interpreted in favor of the ``mixing limited'' picture: for a set amount of turbulence, $\vtL$, one sees a set amount of  cooling, $\edot$, and therefore it is the mixing that determines the dynamics. This interpretation is undercut by the resolution dependence of both cooling and $\edot$ \citep[see e.g.][where resolution dependence of turbulent velocity is shown explicitly in Figure 15]{Lancaster24a}, which calls in to question the overall physical realism of the simulations. The existing results therefore offer limited discriminating power between competing physical interpretations.

%There have also be works in the literature which include explicit viscosity \citep{BS16,TMG25}.
The simulations of \citet{TMG25} explicitly vary $\Pr$ by including explicit viscosity. In the fast-cooling limit, viscosity is observed to increase the total cooling (see their Figures 6 \& 7). Upon inspection of the bottom panels of Figure 5 of \citet{TMG25}, it is our suspicion that these fast-cooling simulations lie in a regime of parameter space close to those simulations discussed in \autoref{app:res_dep}. Specifically, $\Aint(\dx)$ in the simulations is artificially low as proper turbulent folding is not able to be faithfully represented at that resolution. With this in mind, as the viscosity is increased, the momentum dissipation thickens the interface layer, creating a higher volume of gas at intermediate temperatures and hence more cooling. In this way, viscosity mediates an effective increase in $\vdiff$ and $\edot$ increases, as is observed. To be explicit: this is our speculation and we cannot confirm this independently with the results presented here.

In the slow-cooling limit of the \citet{TMG25} work $\edot$ remains relatively unchanged as the viscosity is increased. This is surprising especially because, unlike the fast-cooling regime, it is apparent from their Figures 5 \& 8 that these simulations begin with well-resolved turbulent mixing at low viscosity (an apparent ``mixing limited'' regime) and transition to a nearly turbulence-free, smooth interface (apparently ``diffusion limited'') at high viscosity. Additionally, the diffusion in the latter situation is mediated by viscosity, but it is the thermal diffusivity which should matter for $\edot$ in these simulations. The thermal diffusivity in these simulations is still fundamentally numerical, but its nature is likely strongly affected by the presence of the strong viscosity. This viscosity-independence seems to point towards some convergent dynamics between the resolved viscosity $\nu$ and the numerical thermal diffusivity $\anum$ which drives the ``effective'' ${\rm Pr}_{\rm eff} \equiv \nu/\alpha_{\rm num } \to 1$ (at least in the way that $\anum$ maps to $\vdiff$) when we would normally expect ${\rm Pr} > 1$. Again, this is our speculation, truly understanding these dynamics would require a suite of simulations which vary both resolution and other properties that might affect the relationship between $\nu$ and $\anum$ such as the density contrast, $\chi$, and the mach number of the flow, $\mach$. This is an important objective for future studies.

There have also been a limited number of studies which explore varying ${\rm Pr}$ by including explicit conduction \citep{BS16,Tan21,TanOh21}.
%\textbf{The results of \citet{BS16} are difficult to interpret in our present context because (i) they are simulating a wind tunnel-like set up, which is intrinsically different than our present, more-idealized discussion, (ii) they primarily investigate the effect of magnetic fields and anisotropic thermal conduction, and (iii) while they do not explicitly state it, they give enough information for the reader to identify that the Field-length at the peak cooling temperatures in the layer is not well resolved. While this work still presents interesting results on the interplay of (anisotropic-)conduction and magnetic fields on the clouds, its use to our present conversation is limited.}
In \citet{Tan21} the strength of conduction is steadily increased and it is observed that $\edot$ does not change until the conduction begins to affect the integral-scale turbulent structure (see their Figure 16, and discussion in Section 5.5). This observation is used to argue for a ``mixing limited'' picture. This may well be the case, though we would argue that this could be due to the limited dynamical range of the simulations. In particular, if the simulation resolution is such that conduction is only well resolved (by which we mean dominant to numerical thermal diffusivity) once it begins to affect the
%turbulent cascade
integral-scale turbulence, no clear conclusion can be drawn. The resolution required to have a well-developed inertial range in turbulence, along with resolved dissipative physics, could be quite high, potentially much higher than any of the simulations presented here \citep{Kritsuk07}, making a strong case for future work. In light of the resolution dependence of the magnetized case, and the potential resolution requirements discussed above we would like to generally caution that the addition of new physics to mixing layer problems should not appeal to the resolution independence of simulations like the ones presented here in order to justify the lack of a resolution study.

In the simulations presented here, mixing and cooling dominate the phase space evolution of a fluid element. This makes movement through the mixing layer effectively a `one way street' in that, once gas is mixed, it will for the most part cool to lower temperatures. Therefore there is only movement of gas from hot to cold, and the cold phase grows in mass. The inclusion of explicit thermal conductivity provides another mechanism to change the energy content of a fluid element and can therefore change this picture, with strong thermal conduction in the hot phase potentially resulting in mass loss from the cold phase \citep{CowieMcKee77,ElBadry19}, and much more emission from hot gas \citep{TanOh21} that could be much more consistent with observations \citep[e.g.][]{Rodriguez2025}. We leave study of these effects, along with a controlled study of the effects of varying $\Pr$ to future work.

\subsection{Summary of Results}
\label{subsec:summary}

Our main findings:
\begin{itemize}
    \item[1.] We are able to accurately measure the components of the diffusive enthalpy flux into the mixing layer ($\vdiff$ and $\Aint$) and show that these measurements accurately predict total cooling in the simulation (\autoref{subsec:res0} and \autoref{fig:edot_vs_hflux}), as expected.

    \item[2.] Investigating the behavior of each of these quantities with resolution, $\nres$, we find that apparent resolution independence of total cooling in these mixing layers (\autoref{fig:edot_xi}) is dependent upon the countervailing effects of decreased numerical dissipation (lower $\vdiff$) and decreased numerical viscosity (higher $\Aint$) as shown in \autoref{fig:scaling_cancel}. The behavior of $\vdiff$ is explained through the expected scaling for numerical dissipation (\autoref{fig:vdiff_numeric}) and that of $\Aint$ is explained by an effective fractal model (\autoref{eq:aint_frac}), further explored in \paperii. Whether or not this cancellation can be trusted as physically meaningful is discussed in \autoref{sec:discussion}.

    \item[3.] Despite this cancellation effect, we find that smooth transitions in the the layer (\autoref{fig:teddy_Tsf}) and consequently its phase structure (\autoref{fig:phase_plot}) are only well resolved once $\lcool$ is well resolved. This is the scale on which turbulent diffusion competes on fair terms with cooling (\autoref{sec:theory}).
\end{itemize}
Therefore, we have shown that simulations with $\dx > \lcool$ do not accurately represent turbulent mixing—they capture numerical diffusion across an advected surface. These are not true mixing layers; they are representations of mixing layers, with the representation's limitations hidden by a cancellation of competing terms whose physical realism is as yet ambiguous.

\acknowledgments

The authors would like to thank Eliot Quataert, Eve C. Ostriker, Brent Tan, S. Peng Oh, Max Gr\"{o}nke, Chang-Goo Kim, Amiel Sternberg, Shyam H. Menon, and Alexander Mayer for useful discussions which improved this work. The authors would like to thank the anonymous referee for their careful reading of the text and thoughtful feedback which improved the quality of this work.

The authors gratefully acknowledge the support of the Kavli Institute for Theoretical Physics's 2024 program on ``Turbulence in Astrophysical Environments,'' where this work was initially conceived, and that of the Aspen Center for Physics's program ``Toward a Holistic Understanding of the Multi-scale, Multiphase Circumgalactic Medium'' where this work was continued. This research was therefore supported in part by grant NSF PHY-2309135 to the Kavli Institute for Theoretical Physics and NSF PHY-2210452 to the Aspen Center for Physics. L.L. acknowledges the support of the Simons Foundation under grant 965367. R.M. is supported by National Science Foundation (NSF) grants AST-2107872 and AST-2509269. D.B.F. gratefully acknowledges support from NSF through grants AST-2407387 and from NASA through grants HST-AR-17859.015-A and HST-AR-17559.009-A. This work was supported by a grant from the Simons Foundation (Grant Award ID BD-Targeted-00017375, DBF). G.L.B. acknowledges support from the NSF (AST-2108470, AST-2307419), NASA TCAN award 80NSSC21K1053, and the Simons Foundation through the Learning the Universe Collaboration.
The simulations presented in this work and much subsequent analysis was performed on the Flatiron Institute's \texttt{rusty} computing cluster.
The analysis presented in this article was performed in part on computational resources managed and supported by Princeton Research Computing, a consortium of groups including the Princeton Institute for Computational Science and Engineering (PICSciE) and the Office of Information Technology's High Performance Computing Center and Visualization Laboratory at Princeton University.
This research used both the DeltaAI advanced computing and data resource, which is supported by the NSF (award OAC 2320345) and the State of Illinois, and the Delta advanced computing and data resource which is supported by the NSF (award OAC 2005572) and the State of Illinois. Delta and DeltaAI are joint efforts of the University of Illinois Urbana-Champaign and its National Center for Supercomputing Applications.
\software{
{\tt scipy} \citep{scipy},
{\tt numpy} \citep{harrisNumpy2020}, 
{\tt matplotlib} \citep{matplotlib_hunter07},
{\tt adstex} (\url{https://github.com/yymao/adstex})
}

\newpage

\appendix

\begin{figure*}
    \centering
    \includegraphics[width=\textwidth]{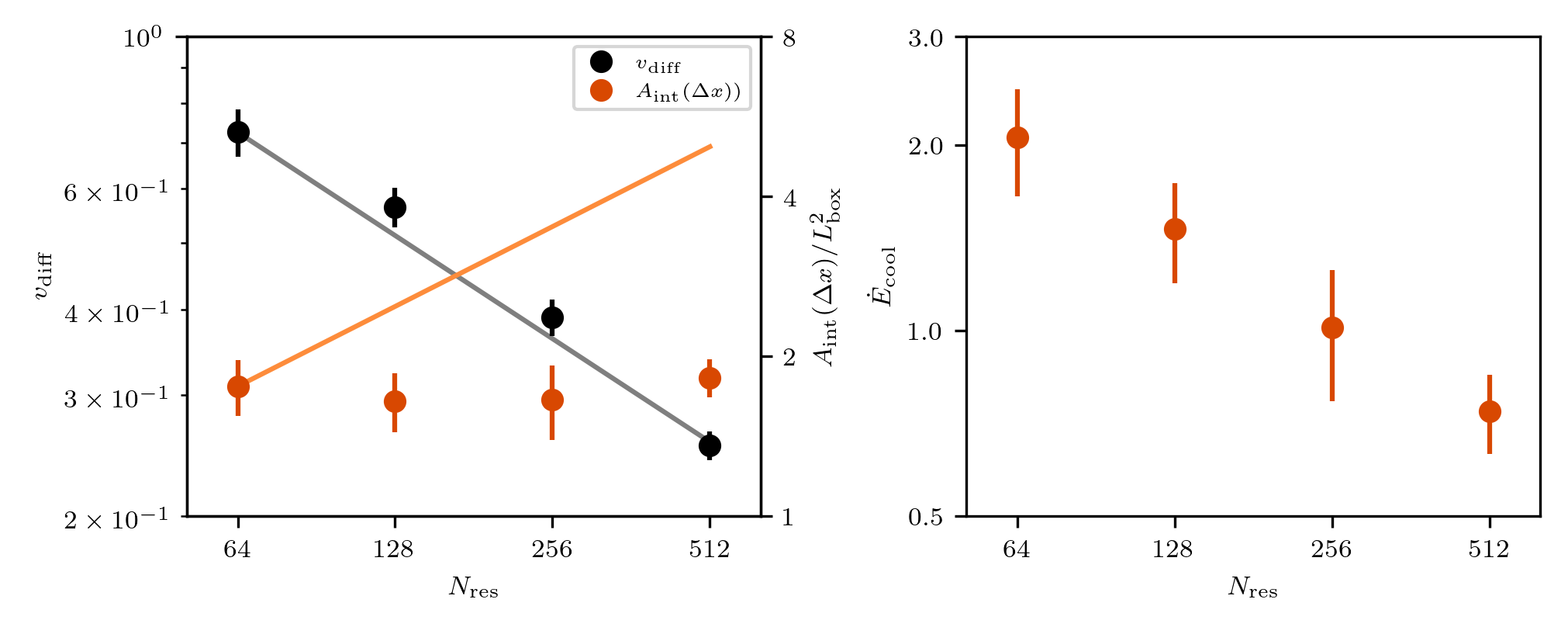}
    \caption{Results for $\chi = 10$, $\xi = 10^3$, and $\mach=1/2$ simulations as a function of resolution. \textit{Left Panel}: A plot analogous to the panels of \autoref{fig:scaling_cancel} showing the behavior of $\vdiff$ (black points, left hand $y$-axis) and $\Aint(\dx)$ (orange points, right hand $y$-axis) as a function of resolution. Scaling proportional to $\propto \nres^{-1/2}$ (black) and  $\propto \nres^{1/2}$ (orange) are shown to guide the eye. \textit{Right Panel}: The total cooling in the simulation as a function of resolution. Total cooling is not resolution independent in these simulations, as $\Aint(\dx)/\Lbox^2 \approx 1$, it's minimum value.}
    \label{fig:low_chi_res}
\end{figure*}

\section{Resolution Dependence at Low Density Contrast}
\label{app:res_dep}

Here we show results from one of our simulations run at low density contrast, specifically $\chi = 10$, $\xi =10^3$, and $\mach=1/2$. As we explain further in \paperii\ in the context of other simulations, in the $\xi \gg 1$ simulations the ram-pressure of the inflowing gas suppresses the turbulent folding of the interface so that $\Aint(\dx)/\Lbox^2 \approx 1$, its minimum allowed value. This remains the case as the simulations move to higher resolution, as we can see in the orange points in the left hand panel of \autoref{fig:low_chi_res}. Since the diffusive velocity continues to decrease at higher resolution (black points in left hand panel of \autoref{fig:low_chi_res}), as we would expect based on the arguments of \autoref{subsec:cooling_and_hflux}, this results in resolution dependence in the total cooling, as seen in the right hand panel of \autoref{fig:low_chi_res}. In this regime, which may well be the regime that many global simulations lie in, turbulence is not well-resolved enough to excite any semblance of complex structure in the interface between the hot and cold gas. In this limit, at these resolutions, the turbulent folding of the interface that creates this structure is inhibited by the dynamical pressure of the cooling-induced inflow to the surface. At high enough resolution, we would expect this folding to eventually become resolved, and for the cancellation of $\Aint(\dx)$ and $\vdiff$ to resume. The effect of the inflow suppressing turbulent folding is the main focus of \paperii.

\begin{figure*}
    \centering
    \includegraphics[width=\textwidth]{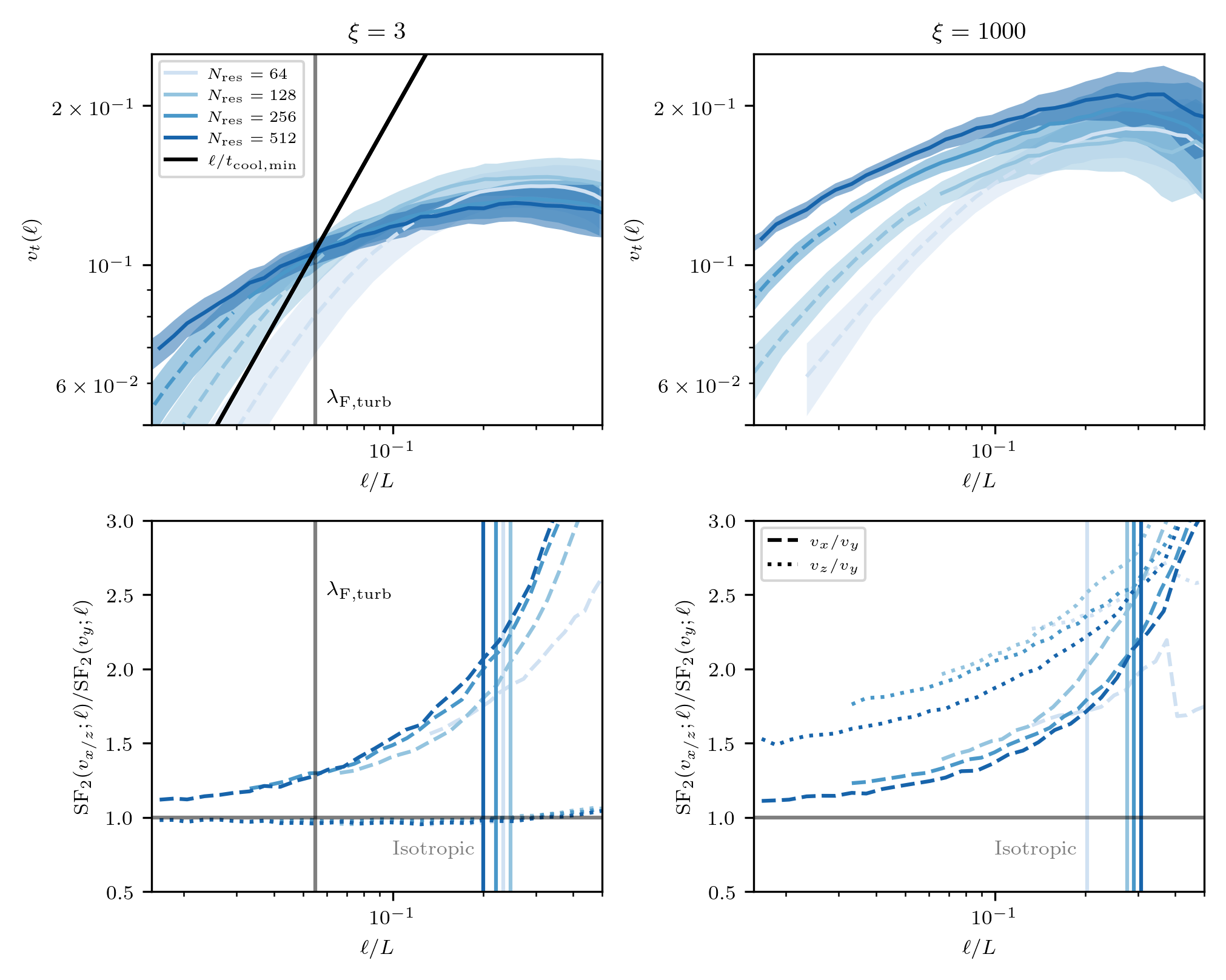}
    \caption{The 2nd order velocity SFs as a function of scale in the $\xi = 3$ (left panel) and $\xi =1000$ (right panel) simulations. \textit{Top panels}: The velocity structure functions $\vtl$ as defined by \autoref{eq:vsf2_def} and measured as described in \autoref{subsec:sf_measure}. We show the velocity scale $\ell/\tcoolmin$ in the $\xi=3$ simulations to indicate the scale $\lcool$. \textit{Bottom panels}: Ratios of structure functions in different directions, ${\rm SF}_2(v_x;\ell)/{\rm SF}_2(v_y;\ell)$ (dashed lines) and ${\rm SF}_2(v_z;\ell)/{\rm SF}_2(v_y;\ell)$ (dotted lines) (${\rm SF}_2(\cdot;\ell)$ defined in \autoref{eq:vsf_vi}). Vertical colored lines indicate the approximate scale height of the mixing layer, $h_{\rm sh}$ (\autoref{eq:hsh_def}). For the bottom panels we restrict to scales $\ell > 8\,\dx$. Darker curves are higher resolution simulations. The vertical gray line in the left panels shows $\lcool$ as in \autoref{fig:teddy_Tsf}.}
    \label{fig:vsfs_isotropy}
\end{figure*}

\section{Turbulent Structure and Isotropy}
\label{app:isotropy}

One of the most tell-tale signs of well-developed hydrodynamical turbulence is that the distribution of velocities is isotropic. In a TRML this can never truly be the case, as the shear direction will always appear as preferred on the largest scales. Nevertheless, in order for our simulations to represent a well-developed turbulent mixing process, it is reasonable to expect that the turbulence appear isotropic on small scales. In order to investigate whether or not this is the case in our simulations we investigate the properties of the turbulent structure functions in \autoref{fig:vsfs_isotropy} for the $\mach=1/2$, $\xi = 3,\ 100$ simulations.

In the top panels of \autoref{fig:vsfs_isotropy} we show the structure functions, measured as described in \autoref{subsec:sf_measure}. Scales $\ell < 8\, \dx$ are indicated by dashed lines in these panels as these are the scales on which it is apparent that numerical viscosity acts to damp turbulent motions. The turbulent structure functions are well-resolved at scales $\ell > 8\,\dx$ except for a mild resolution dependence in the fastest cooling, $\xi=10^3$, mixing layers. The top-left panel also shows the velocity scale $\ell/\tcoolmin$ in a velocity-like analogy to the schematic given in \autoref{fig:resolution_schematic} (the equivalent line falls outside the bounds of the plot for the top-right panel, illustrating that these simulations are far from resolving $\lcool$). The point where the $v_t(\ell)$ curves cross this line is the scale $\lcool$ (\autoref{eq:lcool_def}). It is apparent that this scale is a well-resolved quantity in the $\xi = 3$ simulations which resolve that scale.

In the bottom panels of \autoref{fig:vsfs_isotropy} we show the ratio of the SFs of different velocity components (as measured in \autoref{eq:vsf_vi}). If the turbulence is isotropic on scale $\ell$ then we expect ${\rm SF}_2(v_x;\ell) = {\rm SF}_2(v_z;\ell) = {\rm SF}_2(v_y;\ell)$ so that the ratios will be unity. We see that in the slow cooling regime simulation, where $\xi =3$, the turbulence is indeed isotropic on small scales and the $v_z$ SF remains approximately equal to the $v_y$ over all scales. This is not true of the $v_x$ SF, which becomes relatively large on scales $\ell/L > 0.1$ due to the presence of the background shear motion which is apparent on scales comparable to the scale height of the mixing layer. We approximate this scale height using the $z$-profiles of $v_x$ (measurement detailed in \autoref{subsec:misc_measure}) and calculating the difference between the $z$-value where $v_x = \vrel/4$ and where $v_x = -\vrel/4$:
\begin{equation}
    \label{eq:hsh_def}
    h_{\rm sh} = z\left(\left\langle v_x\right\rangle_z = \vrel/4\right) -  z\left(\left\langle v_x\right\rangle_z = -\vrel/4\right) \, .
\end{equation}
This ``shear scale height'' is shown as vertical colored lines (corresponding to each simulation) in the panels of \autoref{fig:vsfs_isotropy}.

Unlike the $\xi = 3$ simulation, the fast cooling regime $\xi = 1000$ simulation is not isotropic on any scale, with ${\rm SF}_2(v_z)$ remaining much larger than its $v_y$ counterpart down to the smallest scales at the highest resolution. On the largest scales ($\ell \gtrsim h_{\rm sh}$) this is representative of the fact that the bulk inflow to the layer in the $z$ direction is larger than the turbulent motion $\vbulk > \vtL$ (see \paperii). The fact that this is still true on the smallest scales shows that the inflow is still dominant over turbulent motions at these smallest scales. In this regime, the simulation is not representative of true turbulent mixing, but rather an inflow to the surface enabled by the bulk advection of the surface around the grid, caused by numerical diffusion (see \autoref{sec:theory} and \autoref{subsec:cooling_and_hflux}).

We finally note that the degree of anisotropy apparent in \autoref{fig:vsfs_isotropy} clearly demonstrates that the SF of the three-dimensional (3D) velocity field should not be used as a tracer for ``turbulent'' motions, as it would be contaminated by the bulk shear and inflow on large scales. This motivates our choice of $\vtl$ given in \autoref{eq:vsf_measure}.

\begin{figure*}
    \centering
    \includegraphics[width=\textwidth]{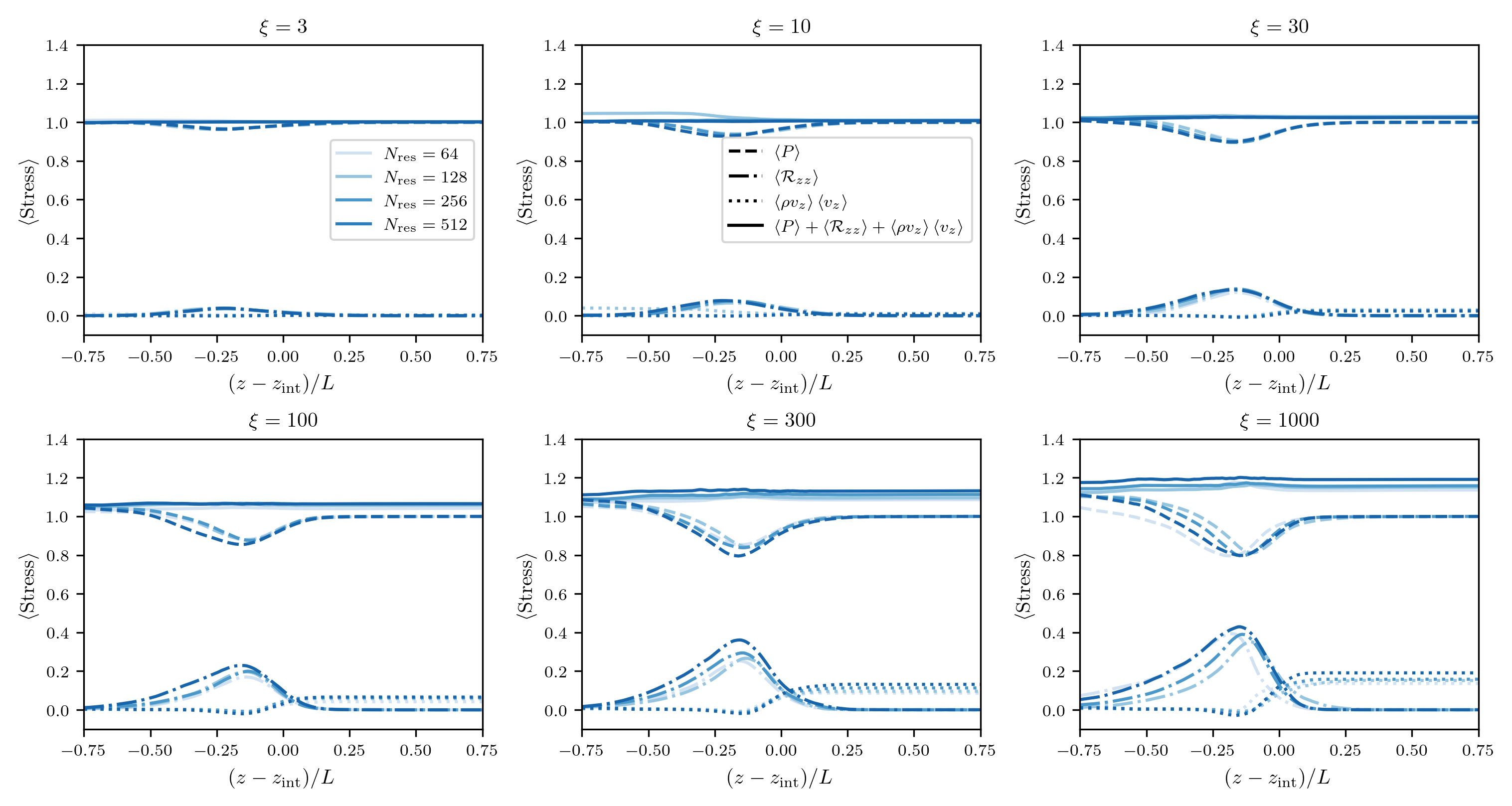}
    \caption{We show the mean-stress profile in the $z$-direction across all simulations with $\mach = 1/2$ at varying resolution and cooling regime. In all simulations (regardless of resolution) pressure, Reynolds stress, and vertical momentum flux balance, as is required for a time-averaged, steady solution to the fluid equations. The relationship between these quantities, however, varies with resolution and regime. This is a recreation of panel (3,2), Figure 2 of \citet{Sharma25} for our simulations.}
    \label{fig:stress_profiles}
\end{figure*}

\section{Reynolds Averaging and Turbulent Pressure Support}
\label{app:turb_support}

As discussed briefly in \autoref{subsec:phase_origin}, \citet{Sharma25} show that well-resolved TRML simulations still have a decrement in the thermal pressure in the interior of the layer and explain that this is allowed through the support of vertical turbulent pressure within the mixing layer. By averaging the $z$-component of the momentum equation over the $x$ and $y$ directions they arrive at \autoref{eq:reynolds_momz}, which states that the averaged pressure, $\left\langle P\right\rangle$, vertical Reynolds stress $\Rzz \equiv \left\langle \delta v_z \delta(\rho v_z) \right\rangle$, and mean compressive stress $\left\langle \rho v_z\right\rangle\left\langle v_z \right\rangle$, balance throughout the layer.

In \autoref{fig:stress_profiles} we show that this relation holds in all of our $\mach = 1/2$ simulations (it also holds in the $\mach = 1/8$ simulations, though we don't show them here). We confirm that turbulent pressure in the layer ($\Rzz > 0$) supports a dip in the mean thermal pressure ($\left\langle P \right\rangle < P_0$). In some of the faster cooling simulations $\Da \gtrsim 50$ we also see that the amount of vertical turbulent stress is not well-converged in some of our simulations. As noted in \autoref{subsec:phase_origin}, this likely explains the over-pressurization (compared to higher resolution simulations) of intermediate temperature gas in these simulations.

Though we do not show it explicitly here or \autoref{fig:phase_plot}, we tested the proposal in \citet{Sharma25} that the volume-weighted temperature PDF of the $x$ and $y$-averaged temperature profile is predicative of the full three-dimensional $T$-PDF (top panels of \autoref{fig:phase_plot}) and did not find good agreement. The averaged temperature profiles are, however, well fit by $\tanh$ profiles in $z$, as proposed in \citet{Sharma25}.


\bibliography{bibliography}{}

@ARTICLE{Weaver77,
       author = {{Weaver}, R. and {McCray}, R. and {Castor}, J. and {Shapiro}, P. and
         {Moore}, R.},
        title = "{Interstellar bubbles. II. Structure and evolution.}",
      journal = {\apj},
         year = 1977,
        month = dec,
       volume = {218},
        pages = {377-395},
          doi = {10.1086/155692},
       adsurl = {https://ui.adsabs.harvard.edu/abs/1977ApJ...218..377W}
}

@ARTICLE{CowieMcKee77,
       author = {{Cowie}, L.~L. and {McKee}, C.~F.},
        title = "{The evaporation of spherical clouds in a hot gas. I. Classical and saturated mass loss rates.}",
      journal = {\apj},
         year = 1977,
        month = jan,
       volume = {211},
        pages = {135-146},
          doi = {10.1086/154911},
       adsurl = {https://ui.adsabs.harvard.edu/abs/1977ApJ...211..135C}
}

@ARTICLE{Stone08_Athena,
       author = {{Stone}, James M. and {Gardiner}, Thomas A. and {Teuben}, Peter and
         {Hawley}, John F. and {Simon}, Jacob B.},
        title = "{Athena: A New Code for Astrophysical MHD}",
      journal = {\apjs},
         year = 2008,
        month = sep,
       volume = {178},
       number = {1},
        pages = {137-177},
          doi = {10.1086/588755},
archivePrefix = {arXiv},
       eprint = {0804.0402},
 primaryClass = {astro-ph},
       adsurl = {https://ui.adsabs.harvard.edu/abs/2008ApJS..178..137S}
}

@ARTICLE{athenapp,
       author = {{Stone}, James M. and {Tomida}, Kengo and {White}, Christopher J. and {Felker}, Kyle G.},
        title = "{The Athena++ Adaptive Mesh Refinement Framework: Design and Magnetohydrodynamic Solvers}",
      journal = {\apjs},
         year = 2020,
        month = jul,
       volume = {249},
       number = {1},
          eid = {4},
        pages = {4},
          doi = {10.3847/1538-4365/ab929b},
archivePrefix = {arXiv},
       eprint = {2005.06651},
 primaryClass = {astro-ph.IM},
       adsurl = {https://ui.adsabs.harvard.edu/abs/2020ApJS..249....4S}
}

@ARTICLE{athenak,
       author = {{Stone}, James M. and {Mullen}, Patrick D. and {Fielding}, Drummond and {Grete}, Philipp and {Guo}, Minghao and {Kempski}, Philipp and {Most}, Elias R. and {White}, Christopher J. and {Wong}, George N.},
        title = "{AthenaK: A Performance-Portable Version of the Athena++ AMR Framework}",
      journal = {arXiv e-prints},
         year = 2024,
        month = sep,
          eid = {arXiv:2409.16053},
        pages = {arXiv:2409.16053},
          doi = {10.48550/arXiv.2409.16053},
archivePrefix = {arXiv},
       eprint = {2409.16053},
 primaryClass = {astro-ph.IM},
       adsurl = {https://ui.adsabs.harvard.edu/abs/2024arXiv240916053S}
}

@ARTICLE{Grete23_num_disp,
       author = {{Grete}, Philipp and {O'Shea}, Brian W. and {Beckwith}, Kris},
        title = "{As a Matter of Dynamical Range - Scale Dependent Energy Dynamics in MHD Turbulence}",
      journal = {\apjl},
         year = 2023,
        month = jan,
       volume = {942},
       number = {2},
          eid = {L34},
        pages = {L34},
          doi = {10.3847/2041-8213/acaea7},
archivePrefix = {arXiv},
       eprint = {2211.09750},
 primaryClass = {astro-ph.GA},
       adsurl = {https://ui.adsabs.harvard.edu/abs/2023ApJ...942L..34G}
}

@ARTICLE{scipy,
       author = {{Virtanen}, Pauli and {Gommers}, Ralf and {Oliphant}, Travis E. and
         {Haberland}, Matt and {Reddy}, Tyler and {Cournapeau}, David and
         {Burovski}, Evgeni and {Peterson}, Pearu and {Weckesser}, Warren and
         {Bright}, Jonathan and {van der Walt}, St{\'e}fan J. and
         {Brett}, Matthew and {Wilson}, Joshua and {Millman}, K. Jarrod and
         {Mayorov}, Nikolay and {Nelson}, Andrew R.~J. and {Jones}, Eric and
         {Kern}, Robert and {Larson}, Eric and {Carey}, C.~J. and
         {Polat}, {\.I}lhan and {Feng}, Yu and {Moore}, Eric W. and {Vand
        erPlas}, Jake and {Laxalde}, Denis and {Perktold}, Josef and
         {Cimrman}, Robert and {Henriksen}, Ian and {Quintero}, E.~A. and
         {Harris}, Charles R. and {Archibald}, Anne M. and
         {Ribeiro}, Ant{\^o}nio H. and {Pedregosa}, Fabian and
         {van Mulbregt}, Paul and {SciPy 1. 0 Contributors}},
        title = "{SciPy 1.0: fundamental algorithms for scientific computing in Python}",
      journal = {Nature Methods},
         year = 2020,
        month = feb,
       volume = {17},
        pages = {261-272},
          doi = {10.1038/s41592-019-0686-2},
archivePrefix = {arXiv},
       eprint = {1907.10121},
 primaryClass = {cs.MS},
       adsurl = {https://ui.adsabs.harvard.edu/abs/2020NatMe..17..261V}
}

@article{scikit-image,
author = {{van der Walt}, St{\'e}fan J. and {Schönberger}, Johannes L. and {Nunez-Iglesias}, Juan and {Boulogne}, François and {Warner}, Joshua D. and {Yager}, Neil and {Gouillart}, Emmanuelle and {Yu}, Tony and the scikit-image contributors},
title = {scikit-image: Image processing in Python},
journal = {PeerJ},
volume = {2:e453},
pages = {1--15},
year = {2014},
doi = {https://doi.org/10.7717/peerj.453},
}

@ARTICLE{harrisNumpy2020,
       author = {{Harris}, Charles R. and {Jarrod Millman}, K. and
         {van der Walt}, St{\'e}fan J. and {Gommers}, Ralf and
         {Virtanen}, Pauli and {Cournapeau}, David and {Wieser}, Eric and
         {Taylor}, Julian and {Berg}, Sebastian and {Smith}, Nathaniel J. and
         {Kern}, Robert and {Picus}, Matti and {Hoyer}, Stephan and
         {van Kerkwijk}, Marten H. and {Brett}, Matthew and {Haldane}, Allan and
         {Fern{\'a}ndez del R{\'\i}o}, Jaime and {Wiebe}, Mark and
         {Peterson}, Pearu and {G{\'e}rard-Marchant}, Pierre and
         {Sheppard}, Kevin and {Reddy}, Tyler and {Weckesser}, Warren and
         {Abbasi}, Hameer and {Gohlke}, Christoph and {Oliphant}, Travis E.},
        title = "{Array Programming with NumPy}",
      journal = {arXiv e-prints},
         year = 2020,
        month = jun,
          eid = {arXiv:2006.10256},
        pages = {arXiv:2006.10256},
archivePrefix = {arXiv},
       eprint = {2006.10256},
 primaryClass = {cs.MS},
       adsurl = {https://ui.adsabs.harvard.edu/abs/2020arXiv200610256H}
}

@article{Lewiner03_marching_cubes,
author = {Thomas Lewiner and Hélio Lopes and Antônio Wilson Vieira and Geovan Tavares and},
title = {Efficient Implementation of Marching Cubes' Cases with Topological Guarantees},
journal = {Journal of Graphics Tools},
volume = {8},
number = {2},
pages = {1--15},
year = {2003},
publisher = {Taylor \& Francis},
doi = {10.1080/10867651.2003.10487582},
URL = {https://doi.org/10.1080/10867651.2003.10487582},
eprint = {https://doi.org/10.1080/10867651.2003.10487582}
}

@article{Chernyaev95,
    author = "Chernyaev, E. V.",
    title = "{Marching Cubes 33: Construction of topologically correct isosurfaces}",
    reportNumber = "CERN-CN-95-17",
    journal = "{GRAPHICON'95}",
    month = "11",
    year = "1995"
}

@inproceedings{lorensen87,
author = {Lorensen, William E. and Cline, Harvey E.},
title = {Marching cubes: A high resolution 3D surface construction algorithm},
year = {1987},
isbn = {0897912276},
publisher = {Association for Computing Machinery},
address = {New York, NY, USA},
url = {https://doi.org/10.1145/37401.37422},
doi = {10.1145/37401.37422},
booktitle = {Proceedings of the 14th Annual Conference on Computer Graphics and Interactive Techniques},
pages = {163–169},
numpages = {7},
series = {SIGGRAPH '87}
}

@ARTICLE{matplotlib_hunter07,
       author = {{Hunter}, John D.},
        title = "{Matplotlib: A 2D Graphics Environment}",
      journal = {Computing in Science and Engineering},
         year = 2007,
        month = may,
       volume = {9},
       number = {3},
        pages = {90-95},
          doi = {10.1109/MCSE.2007.55},
       adsurl = {https://ui.adsabs.harvard.edu/abs/2007CSE.....9...90H}
}

@ARTICLE{ElBadry19,
       author = {{El-Badry}, Kareem and {Ostriker}, Eve C. and {Kim}, Chang-Goo and
         {Quataert}, Eliot and {Weisz}, Daniel R.},
        title = "{Evolution of supernovae-driven superbubbles with conduction and cooling}",
      journal = {\mnras},
         year = 2019,
        month = dec,
       volume = {490},
       number = {2},
        pages = {1961-1990},
          doi = {10.1093/mnras/stz2773},
archivePrefix = {arXiv},
       eprint = {1902.09547},
 primaryClass = {astro-ph.GA},
       adsurl = {https://ui.adsabs.harvard.edu/abs/2019MNRAS.490.1961E}
}

@ARTICLE{Lancaster21a,
       author = {{Lancaster}, Lachlan and {Ostriker}, Eve C. and {Kim}, Jeong-Gyu and {Kim}, Chang-Goo},
        title = "{Efficiently Cooled Stellar Wind Bubbles in Turbulent Clouds. I. Fractal Theory and Application to Star-forming Clouds}",
      journal = {\apj},
         year = 2021,
        month = jun,
       volume = {914},
       number = {2},
          eid = {89},
        pages = {89},
          doi = {10.3847/1538-4357/abf8ab},
archivePrefix = {arXiv},
       eprint = {2104.07691},
 primaryClass = {astro-ph.GA},
       adsurl = {https://ui.adsabs.harvard.edu/abs/2021ApJ...914...89L}
}

@ARTICLE{Lancaster21b,
       author = {{Lancaster}, Lachlan and {Ostriker}, Eve C. and {Kim}, Jeong-Gyu and {Kim}, Chang-Goo},
        title = "{Efficiently Cooled Stellar Wind Bubbles in Turbulent Clouds. II. Validation of Theory with Hydrodynamic Simulations}",
      journal = {\apj},
         year = 2021,
        month = jun,
       volume = {914},
       number = {2},
          eid = {90},
        pages = {90},
          doi = {10.3847/1538-4357/abf8ac},
archivePrefix = {arXiv},
       eprint = {2104.07722},
 primaryClass = {astro-ph.GA},
       adsurl = {https://ui.adsabs.harvard.edu/abs/2021ApJ...914...90L}
}

@ARTICLE{Lancaster24a,
       author = {{Lancaster}, Lachlan and {Ostriker}, Eve C. and {Kim}, Chang-Goo and {Kim}, Jeong-Gyu and {Bryan}, Greg L.},
        title = "{Geometry, Dissipation, Cooling, and the Dynamical Evolution of Wind-blown Bubbles}",
      journal = {\apj},
         year = 2024,
        month = jul,
       volume = {970},
       number = {1},
          eid = {18},
        pages = {18},
          doi = {10.3847/1538-4357/ad47f6},
archivePrefix = {arXiv},
       eprint = {2405.02396},
 primaryClass = {astro-ph.GA},
       adsurl = {https://ui.adsabs.harvard.edu/abs/2024ApJ...970...18L}
}

@ARTICLE{McCourt18,
       author = {{McCourt}, Michael and {Oh}, S. Peng and {O'Leary}, Ryan and {Madigan}, Ann-Marie},
        title = "{A characteristic scale for cold gas}",
      journal = {\mnras},
         year = 2018,
        month = feb,
       volume = {473},
       number = {4},
        pages = {5407-5431},
          doi = {10.1093/mnras/stx2687},
archivePrefix = {arXiv},
       eprint = {1610.01164},
 primaryClass = {astro-ph.GA},
       adsurl = {https://ui.adsabs.harvard.edu/abs/2018MNRAS.473.5407M}
}

@ARTICLE{Ji19,
       author = {{Ji}, Suoqing and {Oh}, S. Peng and {Masterson}, Phillip},
        title = "{Simulations of radiative turbulent mixing layers}",
      journal = {\mnras},
         year = 2019,
        month = jul,
       volume = {487},
       number = {1},
        pages = {737-754},
          doi = {10.1093/mnras/stz1248},
archivePrefix = {arXiv},
       eprint = {1809.09101},
 primaryClass = {astro-ph.GA},
       adsurl = {https://ui.adsabs.harvard.edu/abs/2019MNRAS.487..737J}
}

@ARTICLE{FieldingFractal20,
       author = {{Fielding}, Drummond B. and {Ostriker}, Eve C. and {Bryan}, Greg L. and
         {Jermyn}, Adam S.},
        title = "{Multiphase Gas and the Fractal Nature of Radiative Turbulent Mixing Layers}",
      journal = {\apjl},
         year = 2020,
        month = may,
       volume = {894},
       number = {2},
          eid = {L24},
        pages = {L24},
          doi = {10.3847/2041-8213/ab8d2c},
archivePrefix = {arXiv},
       eprint = {2003.08390},
 primaryClass = {astro-ph.GA},
       adsurl = {https://ui.adsabs.harvard.edu/abs/2020ApJ...894L..24F}
}

@ARTICLE{YangJi23,
       author = {{Yang}, Yanhui and {Ji}, Suoqing},
        title = "{Radiative turbulent mixing layers at high Mach numbers}",
      journal = {\mnras},
         year = 2023,
        month = apr,
       volume = {520},
       number = {2},
        pages = {2148-2162},
          doi = {10.1093/mnras/stad264},
archivePrefix = {arXiv},
       eprint = {2205.15336},
 primaryClass = {astro-ph.GA},
       adsurl = {https://ui.adsabs.harvard.edu/abs/2023MNRAS.520.2148Y}
}

@ARTICLE{CFB23,
       author = {{Chen}, Zirui and {Fielding}, Drummond B. and {Bryan}, Greg L.},
        title = "{The Anatomy of a Turbulent Radiative Mixing Layer: Insights from an Analytic Model with Turbulent Conduction and Viscosity}",
      journal = {\apj},
         year = 2023,
        month = jun,
       volume = {950},
       number = {2},
          eid = {91},
        pages = {91},
          doi = {10.3847/1538-4357/acc73f},
archivePrefix = {arXiv},
       eprint = {2211.01395},
 primaryClass = {astro-ph.GA},
       adsurl = {https://ui.adsabs.harvard.edu/abs/2023ApJ...950...91C}
}

@ARTICLE{FB22,
       author = {{Fielding}, Drummond B. and {Bryan}, Greg L.},
        title = "{The Structure of Multiphase Galactic Winds}",
      journal = {\apj},
         year = 2022,
        month = jan,
       volume = {924},
       number = {2},
          eid = {82},
        pages = {82},
          doi = {10.3847/1538-4357/ac2f41},
archivePrefix = {arXiv},
       eprint = {2108.05355},
 primaryClass = {astro-ph.GA},
       adsurl = {https://ui.adsabs.harvard.edu/abs/2022ApJ...924...82F}
}

@ARTICLE{TonnesenBryan21,
       author = {{Tonnesen}, Stephanie and {Bryan}, Greg L.},
        title = "{It's Cloud's Illusions I Recall: Mixing Drives the Acceleration of Clouds from Ram Pressure Stripped Galaxies}",
      journal = {\apj},
         year = 2021,
        month = apr,
       volume = {911},
       number = {1},
          eid = {68},
        pages = {68},
          doi = {10.3847/1538-4357/abe7e2},
archivePrefix = {arXiv},
       eprint = {2102.05061},
 primaryClass = {astro-ph.GA},
       adsurl = {https://ui.adsabs.harvard.edu/abs/2021ApJ...911...68T}
}

@ARTICLE{Simons2020,
       author = {{Simons}, Raymond C. and {Peeples}, Molly S. and {Tumlinson}, Jason and {O'Shea}, Brian W. and {Smith}, Britton D. and {Corlies}, Lauren and {Lochhaas}, Cassandra and {Zheng}, Yong and {Augustin}, Ramona and {Prasad}, Deovrat and {Snyder}, Gregory F. and {Tollerud}, Erik},
        title = "{Figuring Out Gas \& Galaxies in Enzo (FOGGIE). IV. The Stochasticity of Ram Pressure Stripping in Galactic Halos}",
      journal = {\apj},
         year = 2020,
        month = dec,
       volume = {905},
       number = {2},
          eid = {167},
        pages = {167},
          doi = {10.3847/1538-4357/abc5b8},
archivePrefix = {arXiv},
       eprint = {2004.14394},
 primaryClass = {astro-ph.GA},
       adsurl = {https://ui.adsabs.harvard.edu/abs/2020ApJ...905..167S}
}

@ARTICLE{Abruzzo22,
       author = {{Abruzzo}, Matthew W. and {Bryan}, Greg L. and {Fielding}, Drummond B.},
        title = "{A Simple Model for Mixing and Cooling in Cloud-Wind Interactions}",
      journal = {\apj},
         year = 2022,
        month = feb,
       volume = {925},
       number = {2},
          eid = {199},
        pages = {199},
          doi = {10.3847/1538-4357/ac3c48},
archivePrefix = {arXiv},
       eprint = {2101.10344},
 primaryClass = {astro-ph.GA},
       adsurl = {https://ui.adsabs.harvard.edu/abs/2022ApJ...925..199A}
}

@ARTICLE{Abruzzo24,
       author = {{Abruzzo}, Matthew W. and {Fielding}, Drummond B. and {Bryan}, Greg L.},
        title = "{Taming the TuRMoiL: The Temperature Dependence of Turbulence in Cloud{\textendash}Wind Interactions}",
      journal = {\apj},
         year = 2024,
        month = may,
       volume = {966},
       number = {2},
          eid = {181},
        pages = {181},
          doi = {10.3847/1538-4357/ad1e51},
archivePrefix = {arXiv},
       eprint = {2210.15679},
 primaryClass = {astro-ph.GA},
       adsurl = {https://ui.adsabs.harvard.edu/abs/2024ApJ...966..181A}
}

@ARTICLE{BS16,
       author = {{Br{\"u}ggen}, Marcus and {Scannapieco}, Evan},
        title = "{The Launching of Cold Clouds by Galaxy Outflows. II. The Role of Thermal Conduction}",
      journal = {\apj},
         year = 2016,
        month = may,
       volume = {822},
       number = {1},
          eid = {31},
        pages = {31},
          doi = {10.3847/0004-637X/822/1/31},
archivePrefix = {arXiv},
       eprint = {1602.01843},
 primaryClass = {astro-ph.GA},
       adsurl = {https://ui.adsabs.harvard.edu/abs/2016ApJ...822...31B}
}

@ARTICLE{BSG23,
       author = {{Br{\"u}ggen}, Marcus and {Scannapieco}, Evan and {Grete}, Philipp},
        title = "{The Launching of Cold Clouds by Galaxy Outflows. V. The Role of Anisotropic Thermal Conduction}",
      journal = {\apj},
         year = 2023,
        month = jul,
       volume = {951},
       number = {2},
          eid = {113},
        pages = {113},
          doi = {10.3847/1538-4357/acd63e},
archivePrefix = {arXiv},
       eprint = {2304.09881},
 primaryClass = {astro-ph.GA},
       adsurl = {https://ui.adsabs.harvard.edu/abs/2023ApJ...951..113B}
}

@ARTICLE{ChenOh24,
       author = {{Chen}, Zirui and {Oh}, S. Peng},
        title = "{The survival and entrainment of molecules and dust in galactic winds}",
      journal = {\mnras},
         year = 2024,
        month = jun,
       volume = {530},
       number = {4},
        pages = {4032-4057},
          doi = {10.1093/mnras/stae1113},
archivePrefix = {arXiv},
       eprint = {2311.04275},
 primaryClass = {astro-ph.GA},
       adsurl = {https://ui.adsabs.harvard.edu/abs/2024MNRAS.530.4032C}
}

@ARTICLE{Sharma25,
       author = {{Sharma}, Prateek and {Kumar}, Arnav and {Datta}, Dipayan and {Babul}, Arif and {Das}, Rishita and {Aditya}, Konduri},
        title = "{Universal Structure of Turbulent Radiative Mixing Layers}",
      journal = {arXiv e-prints},
         year = 2025,
        month = sep,
          eid = {arXiv:2509.03802},
        pages = {arXiv:2509.03802},
          doi = {10.48550/arXiv.2509.03802},
archivePrefix = {arXiv},
       eprint = {2509.03802},
 primaryClass = {astro-ph.GA},
       adsurl = {https://ui.adsabs.harvard.edu/abs/2025arXiv250903802S}
}

@ARTICLE{Dutta22,
       author = {{Dutta}, Alankar and {Sharma}, Prateek and {Nelson}, Dylan},
        title = "{Cooling flows around cold clouds in the circumgalactic medium: steady-state models and comparison with TNG50}",
      journal = {\mnras},
         year = 2022,
        month = mar,
       volume = {510},
       number = {3},
        pages = {3561-3574},
          doi = {10.1093/mnras/stab3653},
archivePrefix = {arXiv},
       eprint = {2107.02722},
 primaryClass = {astro-ph.GA},
       adsurl = {https://ui.adsabs.harvard.edu/abs/2022MNRAS.510.3561D}
}

@ARTICLE{Field65,
       author = {{Field}, George B.},
        title = "{Thermal Instability.}",
      journal = {\apj},
         year = 1965,
        month = aug,
       volume = {142},
        pages = {531},
          doi = {10.1086/148317},
       adsurl = {https://ui.adsabs.harvard.edu/abs/1965ApJ...142..531F}
}

@ARTICLE{BegelmanMcKee90,
       author = {{Begelman}, Mitchell C. and {McKee}, Christopher F.},
        title = "{Global Effects of Thermal Conduction on Two-Phase Media}",
      journal = {\apj},
         year = 1990,
        month = aug,
       volume = {358},
        pages = {375},
          doi = {10.1086/168994},
       adsurl = {https://ui.adsabs.harvard.edu/abs/1990ApJ...358..375B}
}

@ARTICLE{ZDPN69,
       author = {{Zel'Dovich}, Ya. B. and {Pikel'Ner}, S.~B.},
        title = "{The Phase Equilibrium and Dynamics of a Gas Volume That is Heated and Cooled}",
      journal = {Soviet Journal of Experimental and Theoretical Physics},
         year = 1969,
        month = jan,
       volume = {29},
        pages = {170},
       adsurl = {https://ui.adsabs.harvard.edu/abs/1969JETP...29..170Z}
}

@ARTICLE{KoyamaInutsuka04,
       author = {{Koyama}, Hiroshi and {Inutsuka}, Shu-ichiro},
        title = "{The Field Condition: A New Constraint on Spatial Resolution in Simulations of the Nonlinear Development of Thermal Instability}",
      journal = {\apjl},
         year = 2004,
        month = feb,
       volume = {602},
       number = {1},
        pages = {L25-L28},
          doi = {10.1086/382478},
archivePrefix = {arXiv},
       eprint = {astro-ph/0302126},
 primaryClass = {astro-ph},
       adsurl = {https://ui.adsabs.harvard.edu/abs/2004ApJ...602L..25K}
}

@ARTICLE{Armillotta17,
       author = {{Armillotta}, L. and {Fraternali}, F. and {Werk}, J.~K. and {Prochaska}, J.~X. and {Marinacci}, F.},
        title = "{The survival of gas clouds in the circumgalactic medium of Milky Way-like galaxies}",
      journal = {\mnras},
         year = 2017,
        month = sep,
       volume = {470},
       number = {1},
        pages = {114-125},
          doi = {10.1093/mnras/stx1239},
archivePrefix = {arXiv},
       eprint = {1608.05416},
 primaryClass = {astro-ph.GA},
       adsurl = {https://ui.adsabs.harvard.edu/abs/2017MNRAS.470..114A}
}

@ARTICLE{GronkeOh18,
       author = {{Gronke}, Max and {Oh}, S. Peng},
        title = "{The growth and entrainment of cold gas in a hot wind}",
      journal = {\mnras},
         year = 2018,
        month = oct,
       volume = {480},
       number = {1},
        pages = {L111-L115},
          doi = {10.1093/mnrasl/sly131},
archivePrefix = {arXiv},
       eprint = {1806.02728},
 primaryClass = {astro-ph.GA},
       adsurl = {https://ui.adsabs.harvard.edu/abs/2018MNRAS.480L.111G}
}

@ARTICLE{LiHopkins20,
       author = {{Li}, Zhihui and {Hopkins}, Philip F. and {Squire}, Jonathan and {Hummels}, Cameron},
        title = "{On the survival of cool clouds in the circumgalactic medium}",
      journal = {\mnras},
         year = 2020,
        month = feb,
       volume = {492},
       number = {2},
        pages = {1841-1854},
          doi = {10.1093/mnras/stz3567},
archivePrefix = {arXiv},
       eprint = {1909.02632},
 primaryClass = {astro-ph.GA},
       adsurl = {https://ui.adsabs.harvard.edu/abs/2020MNRAS.492.1841L}
}

@ARTICLE{Sparre20,
       author = {{Sparre}, Martin and {Pfrommer}, Christoph and {Ehlert}, Kristian},
        title = "{Interaction of a cold cloud with a hot wind: the regimes of cloud growth and destruction and the impact of magnetic fields}",
      journal = {\mnras},
         year = 2020,
        month = dec,
       volume = {499},
       number = {3},
        pages = {4261-4281},
          doi = {10.1093/mnras/staa3177},
archivePrefix = {arXiv},
       eprint = {2008.09118},
 primaryClass = {astro-ph.GA},
       adsurl = {https://ui.adsabs.harvard.edu/abs/2020MNRAS.499.4261S}
}

@ARTICLE{WarrenSchneider25,
       author = {{Warren}, Orlando and {Schneider}, Evan E. and {Mao}, S. Alwin and {Abruzzo}, Matthew W.},
        title = "{Cloud Properties in Simulated Galactic Winds}",
      journal = {\apj},
         year = 2025,
        month = may,
       volume = {984},
       number = {2},
          eid = {191},
        pages = {191},
          doi = {10.3847/1538-4357/adc44c},
archivePrefix = {arXiv},
       eprint = {2410.11747},
 primaryClass = {astro-ph.GA},
       adsurl = {https://ui.adsabs.harvard.edu/abs/2025ApJ...984..191W}
}

@ARTICLE{Mandelker20,
       author = {{Mandelker}, Nir and {Nagai}, Daisuke and {Aung}, Han and {Dekel}, Avishai and {Birnboim}, Yuval and {van den Bosch}, Frank C.},
        title = "{Instability of supersonic cold streams feeding galaxies - IV. Survival of radiatively cooling streams}",
      journal = {\mnras},
         year = 2020,
        month = may,
       volume = {494},
       number = {2},
        pages = {2641-2663},
          doi = {10.1093/mnras/staa812},
archivePrefix = {arXiv},
       eprint = {1910.05344},
 primaryClass = {astro-ph.GA},
       adsurl = {https://ui.adsabs.harvard.edu/abs/2020MNRAS.494.2641M}
}

@ARTICLE{Mohapatra25,
       author = {{Mohapatra}, Rajsekhar and {Dutta}, Alankar and {Sharma}, Prateek},
        title = "{Tracing Multiphase Structure in the Circumgalactic Medium: Insights from Magnetohydrodynamic Turbulence Simulations}",
      journal = {arXiv e-prints},
         year = 2025,
        month = oct,
          eid = {arXiv:2511.00229},
        pages = {arXiv:2511.00229},
          doi = {10.48550/arXiv.2511.00229},
archivePrefix = {arXiv},
       eprint = {2511.00229},
 primaryClass = {astro-ph.GA},
       adsurl = {https://ui.adsabs.harvard.edu/abs/2025arXiv251100229M}
}

@ARTICLE{Mohapatra23,
       author = {{Mohapatra}, Rajsekhar and {Sharma}, Prateek and {Federrath}, Christoph and {Quataert}, Eliot},
        title = "{Multiphase condensation in cluster haloes: interplay of cooling, buoyancy, and mixing}",
      journal = {\mnras},
         year = 2023,
        month = nov,
       volume = {525},
       number = {3},
        pages = {3831-3848},
          doi = {10.1093/mnras/stad2574},
archivePrefix = {arXiv},
       eprint = {2302.09380},
 primaryClass = {astro-ph.GA},
       adsurl = {https://ui.adsabs.harvard.edu/abs/2023MNRAS.525.3831M}
}

@ARTICLE{Tan21,
       author = {{Tan}, Brent and {Oh}, S. Peng and {Gronke}, Max},
        title = "{Radiative mixing layers: insights from turbulent combustion}",
      journal = {\mnras},
         year = 2021,
        month = apr,
       volume = {502},
       number = {3},
        pages = {3179-3199},
          doi = {10.1093/mnras/stab053},
archivePrefix = {arXiv},
       eprint = {2008.12302},
 primaryClass = {astro-ph.GA},
       adsurl = {https://ui.adsabs.harvard.edu/abs/2021MNRAS.502.3179T}
}

@ARTICLE{TanOh21,
       author = {{Tan}, Brent and {Oh}, S. Peng},
        title = "{A model for line absorption and emission from turbulent mixing layers}",
      journal = {\mnras},
         year = 2021,
        month = nov,
       volume = {508},
       number = {1},
        pages = {L37-L42},
          doi = {10.1093/mnrasl/slab100},
archivePrefix = {arXiv},
       eprint = {2105.11496},
 primaryClass = {astro-ph.GA},
       adsurl = {https://ui.adsabs.harvard.edu/abs/2021MNRAS.508L..37T}
}

@ARTICLE{GronkeOh20a,
       author = {{Gronke}, Max and {Oh}, S. Peng},
        title = "{How cold gas continuously entrains mass and momentum from a hot wind}",
      journal = {\mnras},
         year = 2020,
        month = feb,
       volume = {492},
       number = {2},
        pages = {1970-1990},
          doi = {10.1093/mnras/stz3332},
archivePrefix = {arXiv},
       eprint = {1907.04771},
 primaryClass = {astro-ph.GA},
       adsurl = {https://ui.adsabs.harvard.edu/abs/2020MNRAS.492.1970G}
}

@article{damkohler40,
 author = {Damk{\"o}hler, Gerhard},
 journal = {Zeitschrift f{\"u}r Elektrochemie und angewandte
physikalische Chemie},
 number = {11},
 pages = {601--626},
 publisher = {Wiley Online Library},
 title = {Der einfluss der turbulenz auf die
flammengeschwindigkeit in gasgemischen},
 volume = {46},
 year = {1940}
}

@book{kuo12,
 author = {Kuo, Kenneth Kuan-yun and Acharya, Ragini},
 publisher = {John Wiley \& Sons},
 title = {Fundamentals of turbulent and multiphase combustion},
 year = {2012}
}

@book{MoninYaglomBook,
  title={Statistical Fluid Mechanics, Volume II: Mechanics of Turbulence},
  author={Monin, A.S. and Yaglom, A.M.},
  isbn={9780486318141},
  series={Dover Books on Physics},
  url={https://books.google.com/books?id=6xPEAgAAQBAJ},
  year={2013},
  publisher={Dover Publications}
}

@book{PoinsotBook,
  title={Theoretical and Numerical Combustion},
  author={Poinsot, T and Veynante, D.},
  isbn={9781930217102},
  url={https://books.google.com/books/about/Theoretical_and_Numerical_Combustion.html?id=cqFDkeVABYoC},
  year={2005},
  publisher={R.T. Edwards Inc.}
}

@book{PopeTF,
  place={Cambridge},
  title={Turbulent Flows},
  publisher={Cambridge University Press},
  author={Pope, Stephen B.},
  year={2000}}

@book{Frisch95,
  place={Cambridge},
  title={Turbulence: The Legacy of A. N. Kolmogorov},
  publisher={Cambridge University Press},
  author={Frisch, Uriel},
  year={1995}}

@ARTICLE{Kritsuk07,
       author = {{Kritsuk}, Alexei G. and {Norman}, Michael L. and {Padoan}, Paolo and {Wagner}, Rick},
        title = "{The Statistics of Supersonic Isothermal Turbulence}",
      journal = {\apj},
         year = 2007,
        month = aug,
       volume = {665},
       number = {1},
        pages = {416-431},
          doi = {10.1086/519443},
archivePrefix = {arXiv},
       eprint = {0704.3851},
 primaryClass = {astro-ph},
       adsurl = {https://ui.adsabs.harvard.edu/abs/2007ApJ...665..416K}
}

@ARTICLE{Warhaft00,
       author = {{Warhaft}, Z.},
        title = "{Passive Scalars in Turbulent Flows}",
      journal = {Annual Review of Fluid Mechanics},
         year = 2000,
        month = jan,
       volume = {32},
        pages = {203-240},
          doi = {10.1146/annurev.fluid.32.1.203},
       adsurl = {https://ui.adsabs.harvard.edu/abs/2000AnRFM..32..203W}
}

@ARTICLE{Batchelor59,
       author = {{Batchelor}, G.~K.},
        title = "{Small-scale variation of convected quantities like temperature in turbulent fluid. Part 1. General discussion and the case of small conductivity}",
      journal = {Journal of Fluid Mechanics},
         year = 1959,
        month = jan,
       volume = {5},
        pages = {113-133},
          doi = {10.1017/S002211205900009X},
       adsurl = {https://ui.adsabs.harvard.edu/abs/1959JFM.....5..113B}
}

@ARTICLE{CPS91,
       author = {{Constantin}, Petre and {Procaccia}, Itamar and {Sreenivasan}, K.~R.},
        title = "{Fractal geometry of isoscalar surfaces in turbulence: Theory and experiments}",
      journal = {\prl},
         year = 1991,
        month = sep,
       volume = {67},
       number = {13},
        pages = {1739-1742},
          doi = {10.1103/PhysRevLett.67.1739},
       adsurl = {https://ui.adsabs.harvard.edu/abs/1991PhRvL..67.1739C}
}

@ARTICLE{SRM89,
       author = {{Sreenivasan}, K.~R. and {Ramshankar}, R. and {Meneveau}, C.},
        title = "{Mixing, entrainment and fractal dimensions of surfaces in turbulent flows}",
      journal = {Proceedings of the Royal Society of London Series A},
         year = 1989,
        month = jan,
       volume = {421},
       number = {1860},
        pages = {79-107},
          doi = {10.1098/rspa.1989.0004},
       adsurl = {https://ui.adsabs.harvard.edu/abs/1989RSPSA.421...79S}
}

@ARTICLE{Roepke07,
       author = {{R{\"o}pke}, F.~K. and {Hillebrandt}, W. and {Schmidt}, W. and {Niemeyer}, J.~C. and {Blinnikov}, S.~I. and {Mazzali}, P.~A.},
        title = "{A Three-Dimensional Deflagration Model for Type Ia Supernovae Compared with Observations}",
      journal = {\apj},
         year = 2007,
        month = oct,
       volume = {668},
       number = {2},
        pages = {1132-1139},
          doi = {10.1086/521347},
archivePrefix = {arXiv},
       eprint = {0707.1024},
 primaryClass = {astro-ph},
       adsurl = {https://ui.adsabs.harvard.edu/abs/2007ApJ...668.1132R}
}

@ARTICLE{TMG25,
       author = {{Marin-Gilabert}, Tirso and {Gronke}, Max and {Oh}, S. Peng},
        title = "{The (Limited) Effect of Viscosity in Multiphase Turbulent Mixing}",
      journal = {arXiv e-prints},
         year = 2025,
        month = apr,
          eid = {arXiv:2504.15345},
        pages = {arXiv:2504.15345},
          doi = {10.48550/arXiv.2504.15345},
archivePrefix = {arXiv},
       eprint = {2504.15345},
 primaryClass = {astro-ph.GA},
       adsurl = {https://ui.adsabs.harvard.edu/abs/2025arXiv250415345M}
}

@ARTICLE{Das24,
       author = {{Das}, Hitesh Kishore and {Gronke}, Max},
        title = "{Magnetic fields in multiphase turbulence: impact on dynamics and structure}",
      journal = {\mnras},
         year = 2024,
        month = jan,
       volume = {527},
       number = {1},
        pages = {991-1013},
          doi = {10.1093/mnras/stad3125},
archivePrefix = {arXiv},
       eprint = {2307.06411},
 primaryClass = {astro-ph.GA},
       adsurl = {https://ui.adsabs.harvard.edu/abs/2024MNRAS.527..991D}
}

@ARTICLE{ZhaoBai23,
       author = {{Zhao}, Xihui and {Bai}, Xue-Ning},
        title = "{Simulations of weakly magnetized turbulent mixing layers}",
      journal = {\mnras},
         year = 2023,
        month = dec,
       volume = {526},
       number = {3},
        pages = {4245-4261},
          doi = {10.1093/mnras/stad3011},
archivePrefix = {arXiv},
       eprint = {2307.12355},
 primaryClass = {astro-ph.GA},
       adsurl = {https://ui.adsabs.harvard.edu/abs/2023MNRAS.526.4245Z}
}

@ARTICLE{Rodriguez2025,
       author = {{Rodriguez}, Jennifer A. and {Lopez}, Laura A. and {Lancaster}, Lachlan and {Rosen}, Anna L. and {Nayak}, Omnarayani and {Lopez}, Sebastian and {Holland-Ashford}, Tyler and {Webb}, Trinity L.},
        title = "{Taming the Tarantula: How Stellar Wind Feedback Shapes Gas and Dust in 30 Doradus}",
      journal = {arXiv e-prints},
         year = 2025,
        month = dec,
          eid = {arXiv:2512.03129},
        pages = {arXiv:2512.03129},
          doi = {10.48550/arXiv.2512.03129},
archivePrefix = {arXiv},
       eprint = {2512.03129},
 primaryClass = {astro-ph.HE},
       adsurl = {https://ui.adsabs.harvard.edu/abs/2025arXiv251203129R}
}

@ARTICLE{Werk14,
       author = {{Werk}, Jessica K. and {Prochaska}, J. Xavier and {Tumlinson}, Jason and {Peeples}, Molly S. and {Tripp}, Todd M. and {Fox}, Andrew J. and {Lehner}, Nicolas and {Thom}, Christopher and {O'Meara}, John M. and {Ford}, Amanda Brady and {Bordoloi}, Rongmon and {Katz}, Neal and {Tejos}, Nicolas and {Oppenheimer}, Benjamin D. and {Dav{\'e}}, Romeel and {Weinberg}, David H.},
        title = "{The COS-Halos Survey: Physical Conditions and Baryonic Mass in the Low-redshift Circumgalactic Medium}",
      journal = {\apj},
         year = 2014,
        month = sep,
       volume = {792},
       number = {1},
          eid = {8},
        pages = {8},
          doi = {10.1088/0004-637X/792/1/8},
archivePrefix = {arXiv},
       eprint = {1403.0947},
 primaryClass = {astro-ph.CO},
       adsurl = {https://ui.adsabs.harvard.edu/abs/2014ApJ...792....8W}
}

@ARTICLE{JenkinsMeloy74,
       author = {{Jenkins}, E.~B. and {Meloy}, D.~A.},
        title = "{A Survey with Copernicus of Interstellar O VI Absorption}",
      journal = {\apjl},
         year = 1974,
        month = nov,
       volume = {193},
        pages = {L121},
          doi = {10.1086/181647},
       adsurl = {https://ui.adsabs.harvard.edu/abs/1974ApJ...193L.121J}
}

@ARTICLE{Zahedy19,
       author = {{Zahedy}, Fakhri S. and {Chen}, Hsiao-Wen and {Johnson}, Sean D. and {Pierce}, Rebecca M. and {Rauch}, Michael and {Huang}, Yun-Hsin and {Weiner}, Benjamin J. and {Gauthier}, Jean-Ren{\'e}},
        title = "{Characterizing circumgalactic gas around massive ellipticals at z {\ensuremath{\sim}} 0.4 - II. Physical properties and elemental abundances}",
      journal = {\mnras},
         year = 2019,
        month = apr,
       volume = {484},
       number = {2},
        pages = {2257-2280},
          doi = {10.1093/mnras/sty3482},
archivePrefix = {arXiv},
       eprint = {1809.05115},
 primaryClass = {astro-ph.GA},
       adsurl = {https://ui.adsabs.harvard.edu/abs/2019MNRAS.484.2257Z}
}

@ARTICLE{KK13,
       author = {{Kim}, Jeong-Gyu and {Kim}, Woong-Tae},
        title = "{Instability of Evaporation Fronts in the Interstellar Medium}",
      journal = {\apj},
         year = 2013,
        month = dec,
       volume = {779},
       number = {1},
          eid = {48},
        pages = {48},
          doi = {10.1088/0004-637X/779/1/48},
archivePrefix = {arXiv},
       eprint = {1310.2940},
 primaryClass = {astro-ph.GA},
       adsurl = {https://ui.adsabs.harvard.edu/abs/2013ApJ...779...48K}
}

@ARTICLE{JenningsLi21,
       author = {{Jennings}, R. Michael and {Li}, Yuan},
        title = "{Thermal instability and multiphase gas in the simulated interstellar medium with conduction, viscosity, and magnetic fields}",
      journal = {\mnras},
         year = 2021,
        month = aug,
       volume = {505},
       number = {4},
        pages = {5238-5252},
          doi = {10.1093/mnras/stab1607},
archivePrefix = {arXiv},
       eprint = {2012.05252},
 primaryClass = {astro-ph.GA},
       adsurl = {https://ui.adsabs.harvard.edu/abs/2021MNRAS.505.5238J}
}

@ARTICLE{Pittard22Winds,
       author = {{Pittard}, J.~M.},
        title = "{Momentum and energy injection by a wind-blown bubble into an inhomogeneous interstellar medium}",
      journal = {\mnras},
         year = 2022,
        month = sep,
       volume = {515},
       number = {2},
        pages = {1815-1829},
          doi = {10.1093/mnras/stac1954},
archivePrefix = {arXiv},
       eprint = {2207.03370},
 primaryClass = {astro-ph.GA},
       adsurl = {https://ui.adsabs.harvard.edu/abs/2022MNRAS.515.1815P}
}

@BOOK{Drainebook,
       author = {{Draine}, Bruce T.},
        title = "{Physics of the Interstellar and Intergalactic Medium}",
         year = 2011,
       adsurl = {https://ui.adsabs.harvard.edu/abs/2011piim.book.....D}
}

@ARTICLE{HLLC94,
       author = {{Toro}, E.~F. and {Spruce}, M. and {Speares}, W.},
        title = "{Restoration of the contact surface in the HLL-Riemann solver}",
      journal = {Shock Waves},
         year = 1994,
        month = jul,
       volume = {4},
       number = {1},
        pages = {25-34},
          doi = {10.1007/BF01414629},
       adsurl = {https://ui.adsabs.harvard.edu/abs/1994ShWav...4...25T}
}

@BOOK{ToroRiemann,
    author = {Toro, Eleuterio},
    year = {2009},
    month = {01},
    pages = {},
    title = {Riemann Solvers and Numerical Methods for Fluid Dynamics: A Practical Introduction},
    journal = {Riemann Solvers and Numerical Methods for Fluid Dynamics},
    doi = {10.1007/b79761}
}

@BOOK{NumericalRecipes,
       author = {{Press}, William H. and {Teukolsky}, Saul A. and {Vetterling}, William T. and {Flannery}, Brian P.},
        title = "{Numerical recipes in C++ : the art of scientific computing}",
         year = 2002,
       adsurl = {https://ui.adsabs.harvard.edu/abs/2002nrca.book.....P}
}
\bibliographystyle{aasjournal}

\end{document}